\documentstyle[aps]{revtex}
\begin{document}
\draft
\title{Shapes and $\beta -$decay in proton rich Ge, Se, Kr and Sr isotopes}
\author{P. Sarriguren, E. Moya de Guerra, and A. Escuderos}
\address{Instituto de Estructura de la Materia,\\
Consejo Superior de Investigaciones Cient\'{\i }ficas, \\
Serrano 123, E-28006 Madrid, Spain}
\maketitle

\begin{abstract}
We study ground states and $\beta -$decay properties of the proton rich
isotope chains Ge, Se, Kr, and Sr. We use a deformed selfconsistent HF+RPA
approach with density-dependent effective interactions of Skyrme type. We
find that most of the isotopes present two HF minima corresponding to two
different shapes. In addition to static quadrupole moments and other ground
state band properties, we present results for Gamow-Teller strength
distributions, as well as for half-lives and summed strengths. The role of
deformation is particularly emphasized.
\end{abstract}



\baselineskip=2\baselineskip


\vskip 2cm

Corresponding Author: P. Sarriguren

Postal Address: Instituto de Estructura de la Materia, C.S.I.C., Serrano
123, 28006 Madrid, Spain

e-mail address: emsarri@iem.csic.es

telephone: 34 91 561 6800 (ext. 1126)

fax: 34 91 585 5184

\newpage

\section{Introduction}

The field devoted to the study of exotic nuclei is nowadays one of the most
fruitful in Nuclear Physics. Experimental work on nuclei far from stability
is providing a wealth of new information \cite{review} that is a challenge
to theory. Of prime importance is to test the predictions on unstable nuclei
of theoretical models that are trusted for their achievements on stable
nuclei.

The interest on exotic nuclei is manifold. On a first step there is the
intrinsic appeal to know better and better those regions of the nuclear
chart unexplored yet. In addition to that, one has particular interesting
problems still open such as the delimitation of the drip lines, the
appearance of new phenomena, absent in stable nuclei, from which we can
learn new aspects of the nuclear structure, or the decay properties of these
radioactive species that are crucial to understand various phases in the
stellar evolution \cite{astro}. Concerning the last point, nuclear
astrophysics is essential to understand the energy generation, the
nucleosynthesis, and the abundance of elements in stars. Nuclear
astrophysics provides the input (decay properties and cross sections for
nuclear reactions of radioactive nuclei), that are needed to model the late
phases of the stellar life. Since this input cannot be determined
experimentally for the extreme conditions of temperature and density that
hold in the interior of the star, reliable theoretical calculations for
these processes are absolutely necessary.

Reliable predictions of $\beta -$decay strength distributions are also
necessary. These are needed for the calculation of beta decay half-lives as
well as for all kinds of $\beta -$delayed processes like $\beta -$delayed
particle emission or $\beta -$delayed fission. The strength distribution
depends on the microscopic structure of the initial and final nuclear wave
functions as well as on the interaction which mediates the decay, it can be
used to infer information on the nuclear structure or to test different
models or approximations. A reliable description of the ground state of the
parent nucleus and of the states populated in the daughter nucleus is
necessary to obtain a good description of the $\beta -$strength
distribution, and vice versa, failures to describe such distributions would
indicate that an improvement of the theoretical formalism is needed.

Among the microscopic nuclear models designed to describe the properties of
the nuclear excitations we can distinguish basically two types of
approaches. 1) One is a phenomenological approach where one takes an
empirical mean field and assumes a simple separable residual interaction. In
this case there is a severe constraint of the method when applied to exotic
nuclei, connected to the empirical choice of the potential well and residual
force. Since such models are based on parameters locally fitted to the
available data on stable nuclei, their extrapolation to exotic nuclei is at
least questionable. 2) The other approach is the selfconsistent approach.
Here the consistency of the picture is stressed using an effective
interaction, usually a Skyrme interaction, that describes successfully the
ground state properties of the nuclei along the periodic table within a
Hartree-Fock calculation, and it is also able to describe the excited states
from an RPA\ calculation with residual interactions obtained from the same
force. The main difficulty is that the complexity of the calculation
increases rapidly with the size of the configuration space and one has to
work within limited spaces for nonseparable forces. The practical advantage
of approach 1) is that it is possible to calculate nuclear excitations in
very large configuration spaces since there is no need to diagonalize
matrices whose dimensions grow with the size of the configuration space \cite
{giai98}.

One way to combine the good features of both approaches is to construct
first the quasiparticle basis selfconsistently from a Hartree-Fock
calculation with density-dependent Skyrme forces and pairing correlations in
BCS, and then to solve the RPA (or QRPA)\ equations with a separable
residual interaction derived from the same Skyrme force. The separable
residual interaction is obtained from the exact particle-hole residual
interaction corresponding to the Skyrme force after averaging over the
nuclear volume. In this way the consistency (mean field and residual
interaction determined from the same effective interaction) and the
manageability (the size of the RPA problem does not increase with increasing
configuration space), are both exploited. One preserves the reliability of a
selfconsistent treatment without loosing the capability of using large
configuration spaces. This is the framework where our calculations are done.

Our procedure can be viewed as an approximation to the method recently
proposed by Van Giai et al. \cite{giai98}. In Ref. \cite{giai98} the exact
particle-hole residual interaction is first reduced to its Landau-Migdal
form and then the RPA\ matrix is expanded into a finite sum of $n$ separable
terms.

In a previous paper \cite{sarri98} we already applied this method to $^{74}$%
Kr with the aim of identifying those elements of the theory to which $\beta- 
$decay may be particularly sensitive. We found that the Gamow Teller (GT)
strength distribution was especially sensitive to the nuclear shape and RPA
correlations, and we also noted the important role played by the two-body
effective interaction, as well as by pairing correlations. Therefore, it was
concluded that deformation, pairing and the RPA treatment are ingredients
that one cannot avoid in a description of the $\beta -$decay in this mass
region. In this paper we use this knowledge to calculate the decay
properties of a series of isotopes that are being presently measured or are
considered as candidates for experimental studies \cite{Isolde}. They are
proton rich nuclei in the mass region around A=70 ($^{64,66,68,70}$Ge, $%
^{68,70,72,74}$Se, $^{72,74,76,78}$Kr, $^{76,78,80,82}$Sr), where
deformation including shape coexistence plays an important role.

The study of these isotope chains is worth for several reasons. First of
all, this mass region is characterized by a very rich structure giving rise
to a large variety of coexistent nuclear shapes. Thus, this region is a good
laboratory to test nuclear structure models. In addition to that, the study
of various isotope chains opens the possibility to distinguish what is
general and what is particular in the behaviour of these nuclei. The
systematics also allows one to observe whether the agreement with experiment
breaks down as we approach the $N=Z$ isotopes ($^{64}$Ge, $^{68}$Se, $^{72}$%
Kr, $^{76}$Sr), that are expected to have some peculiarities because of the $%
T=0$ pairing correlations \cite{nppairing}. Another interesting point to
discuss is whether the strength distributions of $\beta -$decay can be used
to extract information on the nuclear shape since clear differences in these
distributions could appear depending on the shape of the parent nucleus. It
would be interesting to find the most favorable cases for this purpose.
Finally, since this mass region is at the border or beyond the scope of the
full shell model calculations, predictions for the strength distributions,
half-lives, and summed strengths in this mass region obtained from
selfconsistent mean field approaches are of especial relevance since they
will be probably the most reliable calculations. These results could be used
to guide the experimental searches and to compare with other kind of
calculations when available.

The paper is organized as follows. In Section 2 we remind briefly the main
aspects of our approach and establish our choice for the force and pairing
gap parameters. In Section 3 we present the results obtained for the energy
distribution of the Gamow Teller strength in those isotopes, as well as
integrated quantities that are especially relevant because they can be
measured directly such as half-lives or total Gamow Teller strength
contained within the $Q_{EC\text{ }}$ energy window, which in these proton
rich nuclei is quite large. Finally in Section 4 we point out some final
conclusions and remarks.

\section{Summary of the Theory}

In this section we summarize briefly the theory involved in the microscopic
calculations presented in the next Sections. More details can be found in
Refs. \cite{sarri98,predeal}. Our method consists in a selfconsistent
formalism based on a deformed Hartree-Fock (HF) mean field obtained with a
Skyrme interaction including pairing correlations in the BCS approximation.
The single particle energies, wave functions and occupations are generated
from this mean field. We add to the mean field a spin-isospin residual
interaction with a coupling strength derived by averaging over the nuclear
volume the Landau-Migdal force, obtained from the same energy density
functional (and Skyrme interaction) as the HF equation. The residual force
is therefore consistent with the mean field. The equations of motion are
solved in the proton-neutron quasiparticle random phase approximation (QRPA) 
\cite{qrpa}.

The merits of the density-dependent HF approximation to describe the
ground-state properties of both spherical and deformed nuclei are well known 
\cite{flocard}. We consider in this paper the force SG2 \cite{giai} of Van
Giai and Sagawa although we also show results in some instances obtained
with the most traditional Skyrme force Sk3 \cite{beiner}. We use Sk3 in its
density dependent two-body version that has better spin-isospin properties
than the three-body one \cite{sarr}. The two forces were designed to fit
ground state properties of spherical nuclei and nuclear matter properties
but, in addition, the force SG2 gives a good description of Gamow Teller
excitations in spherical nuclei \cite{giai}. It also provides a good
description of spin excitations in deformed nuclei \cite{sarr}.

For the solution of the HF equations we follow the McMaster procedure that
is based in the formalism developed in Ref.\cite{vautherin} as described in
Ref.\cite{vallieres}. The single-particle wave functions are expanded in
terms of the eigenstates of an axially symmetric harmonic oscillator in
cylindrical coordinates. We use eleven major shells. The method also
includes pairing between like nucleons in the BCS approximation with fixed
gap parameters for protons $\Delta _{p},$ and neutrons $\Delta _{n}$, which
are determined phenomenologically from the odd-even mass differences through
a symmetric five term formula involving the experimental binding energies 
\cite{audi}.

In the next Section we will discuss shape coexistence. To that end we
perform constrained HF calculations with a quadratic quadrupole constraint 
\cite{constraint}, and analyze the energy surfaces as a function of
deformation. The curves are obtained by minimizing the HF energy under the
constraint of holding the nuclear deformation fixed. This is carried out
over a large range of deformations. When more than one local minimum occurs
for the total energy as a function of deformation, shape coexistence results.

For the study of $\beta -$decay the relevant residual interactions are the
spin-isospin contact forces generating the allowed Gamow Teller transitions.
Following Bertsch and Tsai \cite{bertsch} the particle-hole interaction
consistent with the HF mean field is obtained as the second derivative of
the energy density functional with respect to the one-body density.
Neglecting momentum dependent terms, this gives a local interaction that can
be put in the Landau-Migdal form \cite{migdal}

\begin{equation}
V_{ph}=N_{0}^{-1}\sum_{\ell =0,1}\left[ F_{\ell }+G_{\ell }\bbox{\sigma_1
\cdot \sigma_2}+\left( F_{\ell }^{\prime }+G_{\ell }^{\prime }\bbox{\sigma_1
\cdot \sigma_2}\right) \bbox{\tau_1 \cdot \tau_2}\right] \delta \left( \text{%
{\bf r}}_{1}-\text{{\bf r}}_{2}\right) .
\end{equation}
Retaining only the $\ell =0$ spin-isospin term and averaging the contact
interaction over the nuclear volume, we end up with a separable residual ph
interaction

\begin{equation}
V_{GT}=2\chi _{GT}\sum_{K}\left( -1\right) ^{K}\;\beta _{K}^{+}\beta
_{-K}^{-}  \label{vgt}
\end{equation}
in terms of the Gamow Teller operator $\beta _{K}^{\pm }=\sigma _{K}t^{\pm
}\;\left( K=0,\pm 1\right) .$ The coupling strength is given by

\begin{equation}
\chi _{GT}=\frac{3}{4\pi R^{3}}\left( -\frac{1}{2}\right) \left\{ t_{0}+%
\frac{1}{2}k_{F}^{2}\left( t_{1}-t_{2}\right) +\frac{1}{6}t_{3}\rho ^{\alpha
}\right\} =N_{0}^{-1}\frac{3G_{0}^{\prime }}{2\pi R^{3}}  \label{chigt}
\end{equation}
as a function of the Skyrme parameters $t_{0},t_{1},t_{2},t_{3},\alpha ,$
the nuclear radius $R,$ and the Fermi momentum $k_{F\text{ }}.$

The proton-neutron QRPA phonon operator for Gamow Teller excitations in
even-even nuclei is written as

\begin{equation}
\Gamma _{\omega _{K}}^{+}=\sum_{pn}\left[ X_{pn}^{\omega _{K}}\alpha
_{n}^{+}\alpha _{\bar{p}}^{+}-Y_{pn}^{\omega _{K}}\alpha _{\bar{n}}\alpha
_{p}\right]  \label{phon}
\end{equation}
where $\alpha ^{+}\left( \alpha \right) $ are quasiparticle creation
(annihilation) operators, $\omega _{K}$ are the excitation energies, and $%
X_{pn}^{\omega _{K}},Y_{pn}^{\omega _{K}}$ the forward and backward
amplitudes, respectively. The advantages of using a separable residual
interaction are well known, the RPA\ problem can be easily solved no matter
how many two-quasiparticle (2qp) configurations are included. The RPA\
eigenvalues are obtained as the root of a single secular equation and then
the corresponding RPA\ amplitudes can be calculated by performing summations
over 2qp states. Explicit expressions of the secular equations that we solve
for the $K=0$ and $K=1$ Gamow Teller modes are given in Ref. \cite{sarri98}.

In the intrinsic frame the Gamow Teller $\beta _{K}^{+}$ strengths
connecting the ground state $0^{+}$ and the excited states $1_{\omega
_{K}}^{+}$ are obtained as

\begin{equation}
\left\langle \omega _{K}\left| \beta _{K}^{+}\right| 0\right\rangle
=\sum_{pn}\left( u_{n}v_{p}X_{pn}^{\omega _{K}}+v_{n}u_{p}Y_{pn}^{\omega
_{K}}\right) \left\langle n\left| \sigma _{K}\right| p\right\rangle
\label{strength}
\end{equation}
where the $v^{\prime }$s are the occupation amplitudes $\left(
u^{2}=1-v^{2}\right) $. From the RPA equations it is easy to go back to
simpler approximations: The Tamm Dancoff approximation (TDA) is recovered by
neglecting all the terms involving the backward amplitudes $Y$. The
uncorrelated two-quasiparticle excitations are obtained in the limit of zero
residual interaction. The Ikeda sum rule is fulfilled in all of these
approximations. For each component $K=0,\pm 1,$ we get

\begin{equation}
\sum_{\omega _{K}}\left| \left\langle \omega _{K}\left| \beta
_{K}^{-}\right| 0\right\rangle \right| ^{2}-\left| \left\langle \omega
_{K}\left| \beta _{K}^{+}\right| 0\right\rangle \right| ^{2}=N-Z,
\label{ikeda}
\end{equation}
and summing over $K$, we obtain $3\left( N-Z\right) $ as expected.

In the laboratory frame the transition probability for $\beta ^{+}$ decay
from the $0^{+}$ to a $1_{\omega }^{+}$ state is given by

\begin{equation}
B_{GT}^{+}\left( \omega \right) =\frac{g_{A}^{2}}{4\pi }\left\{ \sum_{\omega
_{0}}\left| \left\langle \omega _{0}\left| \beta _{0}^{+}\right|
0\right\rangle \right| ^{2}\delta \left( \omega -\omega _{0}\right)
+2\sum_{\omega _{1}}\left| \left\langle \omega _{1}\left| \beta
_{1}^{+}\right| 0\right\rangle \right| ^{2}\delta \left( \omega -\omega
_{1}\right) \right\} .  \label{bgt}
\end{equation}

Finally the half-lives are obtained from the $B_{GT}^{+}$ strengths within
the theoretical energy window $Q_{EC}.$

\section{Results}

\subsection{Ground State Properties}

The constrained HF method allows one to get the best solution for each value
of the mass quadrupole $Q_{0}$. In Figs. 1 to 4 we show the HF energy as a
function of the mass quadrupole moment for the two interactions SG2 (solid)
and Sk3 (dashed) in Ge, Se, Kr, and Sr isotopes, respectively. The best HF
solution at each $Q_{0}$ value is obtained by varying the size and
deformation parameters \cite{vautherin} of the deformed harmonic oscillator
basis containing 286 states (plus their time reverse). One should note that
in these figures the origin of the vertical axis varies for each plot but
the unit length (distance between ticks) corresponds always to 1 MeV. Tables
1-4 contain the values of the binding energies obtained in the various cases
from which one can deduce the appropriate vertical scale.

As it is seen in Figs. 1-4 in most cases there are two minima close in
energy, indicating shape coexistence. Fig. 1 for Ge-isotopes shows that the
two solutions are one in the prolate sector and one in the oblate sector in
the four isotopes studied. The two forces agree in their predictions on the
position of the minima with the only exception of $^{70}$Ge, where Sk3
produces a prolate solution at a larger $Q_{0}$ value than SG2. The energies
of the two minima are quite close (less than 1 MeV apart in all cases),
indicating a very favorable case to find shape coexistence in any of these
four Ge-isotopes studied. It is remarkable in this case the similarity among
the four isotopes. Table 1 contains various ground state properties of these
Ge-isotopes for the oblate and prolate solutions of the forces SG2 and Sk3.
The first columns contain the pairing gap parameters for neutrons $\Delta
_{n}$ and protons $\Delta _{p}$ as derived from the experimental masses \cite
{audi}. Besides the Skyrme force, they are the only input parameters in our
calculation. In the next columns we can find the Fermi energies for neutrons 
$\lambda _{n}$ and protons $\lambda _{p}$, the charge radii $r_{C}$, the
charge $\left( Q_{0,p}\right) $ and mass $\left( Q_{0}\right) $ quadrupole
moments, the quadrupole deformations $\beta _{0}$, the values of $%
\left\langle J^{2}\right\rangle $, the cranking moments of inertia ${\cal I}%
_{cr}$, the gyromagnetic ratios $g_{R}$, the binding energies $E_{T}$, the
coupling constant of the residual interaction $\chi _{GT}$, and the $Q_{EC}$
values.

Fig. 2 and Table 2 are the analogous to Fig. 1 and Table 1 for Se-isotopes.
In this case, we can see from Fig. 2 the existence again of two solutions in
each isotope but now there is a tendency to favor the oblate solution as the
ground state. This is true in the four isotopes and with the two forces
considered. It is also worth mentioning that with the force SG2 the prolate
solution tends to disappear as the number of neutrons increases. When one
reaches $^{74}$Se, only an oblate and a spherical solution survive. Fig. 3
and Table 3 contain the results for the Kr-isotopes. Here, we still find
shape isomerism but the situation now changes considerably from one isotope
to another as well as from one force to another. $^{72}$Kr exhibits a
pronounced oblate ground state shape and a prolate isomer with both forces
SG2 and Sk3. The next isotope $^{74}$Kr exhibits shape isomerism as well,
but its characteristics depend on the force. While SG2 favors an oblate
shape, Sk3 favors a prolate one. The situation changes again in $^{76}$Kr,
where SG2 clearly indicates a spherical ground state while Sk3 predicts an
oblate/prolate coexistence. This is also the case in $^{78}$Kr. Thus, while
the two isomers oblate and prolate survive in all cases with Sk3, the force
SG2 predicts an oblate ground state and a prolate isomer in the $N=Z$
isotope $^{72}$Kr. Little by little the oblate ground state collapses into a
spherical solution as the number of neutrons increases and the prolate
solution finally disappears. Fig. 4 and Table 4 show the results for
Sr-isotopes. We can see in this case that the two forces agree in describing 
$^{82}$Sr and $^{80}$Sr as spherical but they differ in $^{78}$Sr and $^{76}$%
Sr. Sk3 produces a prolate ground state in these two isotopes and a shape
isomer which is oblate but almost spherical. On the other hand, SG2 favors a
spherical ground state in $^{78}$Sr with a prolate isomer and a shape
coexistence oblate/prolate in $^{76}$Sr.

Numerical comparison with experiment of binding energies shows that the SG2
force gives a small overbinding ($\sim 2\%$), while the Sk3 force gives a
small underbinding ($\sim 0.7\%$), systematically in all the nuclei
considered. Consistently, we find that the nuclear size, as represented by
the $r_{C}$ values, are systematically somewhat larger with Sk3 than with
SG2, both being in good agreement with the available experimental values.
This comparison of binding energies and $r_{C}$ values does not point out to
any particular difference between the $N=Z$ and the $N>Z$ even-even isotopes.

Experimental $\left| \beta _{0}\right| $ values, as extracted from $B(E2)$
measurements \cite{raman}, are also in good agreement with most of our
microscopically calculated $\beta _{0}$ values. We would like to recall here
that nonzero experimental $\beta $ values in spherical nuclei correspond to
vibrational excitations rather than to the stable deformations calculated
here, thus the experimental $\left| \beta _{0}\right| $ value for $^{82}$Sr
in table 4, corresponds to a vibrational $E2$ transition. The moments of
inertia and collective gyromagnetic ratios are given for possible future
comparison to theory and experiment. The coupling strengths $\chi _{GT}$ are
obtained from Eq.(\ref{chigt}) using the Sk3 and SG2 Skyrme parameters and $%
R=1.2A^{1/3}$ fm. The $Q_{EC}$ values in Tables 1-4 are calculated from our
theoretical binding energies,

\begin{equation}
Q_{EC}=m_{p}-m_{n}+m_{e}-\left( \lambda _{n}+E_{n}\right) _{\left(
N,Z\right) }+\left( \lambda _{p}-E_{p}\right) _{\left( N,Z-2\right) }
\label{qec}
\end{equation}

\subsection{Gamow Teller Strength Distributions}

Before starting to comment the results obtained for the strength
distributions, a discussion concerning the residual interaction is in order.
As we have already mentioned, the coupling strength of our spin-isospin
residual interaction $\chi _{GT\text{ }}$is obtained from the Skyrme
parameters (Eq.(\ref{chigt})) and therefore, the mean field and residual
interaction are consistently derived from the same force without any free
parameter left. Nevertheless, one could ask how this coupling strength
compares with other values previously used in the literature and how well it
describes the position of the experimental Gamow Teller resonance (GTR) as
obtained from $\left( p,n\right) $ reactions. Such comparisons to experiment
have been a common method  to adjust the coupling strength of the residual
spin-isospin force. By this procedure a value of $\chi _{GT}=23/A$ MeV was
obtained \cite{gaarde} for the coupling strength in Eq.(\ref{vgt}). The fit
corresponds to the GTR in $^{208}$Pb, which is centered at an excitation
energy in the daughter nucleus of 15.5 MeV. The value for $\chi _{GT\text{ }}
$ was obtained by using the experimental values for the particle and hole
energies as explained in Ref.\cite{gaarde} and then it would change if one
uses, instead of those experimental energies, the single particle energies
as obtained from a selfconsistent mean field calculation as in our case.
That means that one should be careful when using this value for the coupling
strength because it also implies the use of the experimental energies. As
soon as one uses a different set of single particle energies, the fitting
procedure should be repeated to extract a new value of $\chi _{GT}$ able to
reproduce the excitation energy of the GTR within the new framework.

It should also be mentioned that the value of the coupling strength that
reproduces the position of the GTR in $^{208}$Pb varies if one considers a
different mass region. It is known \cite{homma} that one needs different
values of $\chi _{GT}$ to reproduce the GTR in different mass regions. In an
attempt to improve the systematics of the dependence of the strength $\chi
_{GT}$ with the mass number $A$, more sophisticated dependences than $%
\kappa /A$ have been tried \cite{homma}. A dependence of the type $\chi
_{GT}=\kappa /A^{\mu }$ has been adjusted to data in Ca, Zr, and Pb. It
has been found that $\chi _{GT}=5.2/A^{0.7}$ is able to reproduce in a
reasonable way those data. Again, this parametrization would be dependent on
the mean field and single particle energies used.

The value we obtain for the coupling strength is 27/A MeV for SG2 and 26/A
MeV for Sk3, which are quite close and a little bit higher than the value
23/A, mentioned above. A similar value, 28/A MeV, was obtained in Ref. \cite
{suzuki} to reproduce the systematics of the energy differences between the
GTR and the isobaric analog state observed in $\left( p,n\right) $
reactions. With our value we obtain for $^{208}$Pb the position of the GT
resonance at 19 MeV which is a few MeV larger than experiment. This result
was already known for the force SG2. In Ref. \cite{giai} it was found,
within a TDA calculation with a contact Landau force in $^{208}$Pb, that SG2
gives the GTR at 18 MeV while the other force considered in that paper (SG1)
produces the peak at 21 MeV. In Ref. \cite{liu} the resonance was also found
at 19 MeV within an RPA calculation with a Skyrme-Landau interaction. Our
results from a separable equivalent force confirm those results. The value
of $\chi _{GT}$ needed to reproduce in our case the GTR in $^{208}$Pb is $%
\chi _{GT}=19/A$ MeV.

To illustrate further the comparison of the calculated and experimental
positions of the GTR, it is interesting to compare the predictions of our
approach using the consistent separable residual interaction in the mass
region of our interest here. Unfortunately there is not much experimental
information available. One exception is the case of the Fe isotopes, which
are probably the most extensively studied nuclei in this mass region because
of the interest in astrophysics. The isotopes $^{54,56}$Fe have been
measured by $\left( p,n\right) $ \cite{rapaport} and $\left( n,p\right) $ 
\cite{elka} reactions to obtain the GT$_{\text{ }}^{-}$ and GT$^{+}$
strengths, respectively. We can see in Fig. 5 the result of this comparison,
where the strength ($L=0$ forward-angle cross section or GT strength) has
been plotted versus the excitation energy of the daughter nucleus. The
agreement with the experimental position of the GTR is quite reasonable,
especially if one takes into account that the theoretical calculation has no
free parameters. It is also clear that one could improve this agreement by
reducing a little bit the coupling strength.

Experimental information on $\left( n,p\right) $ reactions is also available
for $^{70,72}$Ge \cite{vetterli}, which are much closer to the nuclei of our
interest in this paper. Fig. 6 contains this comparison between the
experimental $L=0$ cross sections and the Gamow Teller strength distribution
calculated with the force SG2 in RPA and for the two shapes (prolate and
oblate in $^{70}$Ge, spherical and oblate in $^{72}$Ge) that produce HF
energy minima. As in the case of the Fe isotopes, the agreement with the
experimental excitation energy of the GTR is not bad. The peak in $^{70}$Ge
is well reproduced while the peak in $^{72}$Ge is at the correct energy
although experimentally it appears as a broad resonance.

It is not the aim of this paper to fit experimental data on GT strength
distributions, but rather to provide results obtained from the consistent
value of $\chi _{GT}$ --as obtained from the Skyrme force-- to avoid playing
around with free parameters. In any event, the comparison on Fe and Ge above
discussed, shows that the method gives reasonable results. On the other
hand, the strength below the $Q-$window, which is the relevant energy region
for $\beta -$decay, is in practically all the cases considered here much
below the peak of the GTR and therefore not influenced directly by its
position within a few MeV.

The Gamow Teller $\beta ^{+}$ strength distributions calculated in the
selfconsistent HF+RPA scheme with the force SG2 are shown in Figs. 7-10 for
the Ge, Se, Kr, and Sr-isotopes, respectively, as a function of the
excitation energy of the daughter nucleus. We have folded the calculated GT
strengths with $\Gamma =1$ MeV width Gaussians converting the discrete
spectrum into a continuous curve. In these figures, the GT strength of the
various isotopes are compared among themselves in a different panel for each
nuclear shape. In this way one can appreciate the magnitude of the various
strengths on the same scale.

Fig. 7 shows the GT distributions in Ge-isotopes. If one concentrates first
on the comparison of the strengths for a given shape, the first thing to
notice is that the main peaks of the strength occur at lower energies when
one increases the number of neutrons. This is accompanied with a reduction
of the strength with increasing neutron number. Now, if we compare the
strength distributions of a given isotope obtained from the two shapes, we
find that the profiles are in this case quite similar. They are peaked at
about the same energy and contain comparable strengths, the oblate ones
being a little bit smaller. This is true for the four Ge-isotopes considered
and therefore, we can conclude that the Ge-isotopes are not among the best
candidates to look for deformation effects based on the GT strengths
distributions.

Fig. 8 contains the GT strength distributions for the Se-isotopes. In this
case we also find a clear reduction of the GT strength with the increasing
number of neutrons in the oblate and prolate solutions. However, contrary to
what happened with the Ge-isotopes, we observe now that the position of the
main peaks does not become systematically lower with increasing neutron
number, on the contrary, the energy of the main peaks is quite similar for
all the isotopes except for the $N=Z$ one that is shifted to higher energies
in both cases oblate and prolate. A comparison between the oblate and
prolate strength distribution for a given isotope shows that there are not
substantial differences between them. The position of the peaks appear at
about the same energy and only a slightly smaller strength in the oblate
case is worth mentioning. It should also be mentioned that for $^{74}$Se the
curve shown on the left panel (prolate label) corresponds actually to the
spherical solution in Fig. 2 since there is no prolate solution for this
nucleus with the SG2 force.

Fig. 9 is the analogous for Kr isotopes. Here one should also take into
account that for $^{78}$Kr there is a single spherical solution and that the
results shown in the panels labelled oblate and prolate correspond actually
to this spherical solution. Similarly, the results under the label oblate
for $^{76}$Kr corresponds actually to the spherical solution in Fig. 3. In
this chain of isotopes the changes are more dramatic, especially in what
concerns the oblate and prolate differences. The strength again increases as
we approach the $N=Z$ isotope and the position of the bumps is also
displaced to higher energies. The important new feature here is the strong
difference between the calculated strength distributions obtained for the
two different shapes. The most remarkable differences are those between the
oblate and prolate solutions in $^{74}$Kr and between the prolate and
spherical solutions in $^{76}$Kr. Here we have found firm candidates to
study the shapes from their decay properties. Note that the figures
corresponding to $^{74}$Kr are slightly different from those shown in Ref. 
\cite{sarri98}. This is simply due to the different values of $\chi _{GT}$
used that correspond to a different choice of the nuclear radius $R$ in Eq. (%
\ref{chigt}). In this paper $R=1.2A^{1/3}$fm.

The strength distributions in Sr isotopes can be seen in Fig. 10. The trend
observed within the prolate solutions is similar to the above mentioned
behavior, the strength increases and is shifted to higher energies as we
approach $N=Z$. In the right hand side panel, the strength for $^{76}$Sr
corresponds actually to decay from the oblate solution in Fig. 4. For the
true spherical cases ($^{78,80,82}$Sr), the strengths are noticeably smaller
as compared to the deformed shapes. Therefore, this fact can be exploited to
study nuclear shapes from $\beta -$decay properties as in the previous case
for Kr isotopes. We will come back to this point when discussing the
strengths summed up to the accessible experimental window in the next
subsection.

In the next set of figures, Figs. 11-14, we compare the GT strength
distributions obtained in RPA, TDA, and in the uncorrelated
two-quasiparticle case with the force SG2. The general trend seen in these
figures is similar to that observed in our previous work on $^{74}$Kr \cite
{sarri98} and can be summarized as follows: Compared to the uncorrelated two
quasiparticle response, RPA produces two types of effects. First, there is a
shift of the GT strength to higher energies due to the repulsive character
of the spin-isospin residual interaction and second, there is a reduction of
the total strength. While the shifting effect is already contained in the
TDA description, the quenching effect is not.

Fig. 11 shows this comparison among different approximations for the oblate
and prolate shapes of the Ge isotopes. We can see explicitly on the example
of this figure the two effects just described. The displacement of the
strength to higher excitation energies in TDA and RPA with respect to the
uncorrelated case, and the suppression of the strength in RPA. We can also
study the dependence on deformation of the GT strength distributions in the
uncorrelated basis. If we compare the uncorrelated prolate and oblate
distributions (dotted lines) for a given isotope, we arrive to the same
conclusion as in the discussion of Fig. 7. There is not a strong dependence
on deformation for these Ge isotopes, although now some differences become
more apparent. For example there is a first bump at very small energies in
all the oblate cases that is almost suppressed in the prolate ones. These
bumps are redistributed by the action of the residual force and a much
smoother strength distribution is found in RPA. Nevertheless, there is still
a small bump, reminiscent of the peak in the uncorrelated case, that appears
at small energies in the oblate cases and that, as we shall see later on,
plays an important role because it is a signature of an oblate shape in the
parent nucleus that can be identified by measuring the GT strength at low
excitation energies below the $Q_{EC}$ window. Thus, although the RPA\
strength distributions are smooth out in comparison to the more sensitive
uncorrelated distributions, there are still traces of that sensitivity which
can be exploited to probe the shape of the nucleus.

The effect of residual interaction and RPA correlations for Se, Kr, and Sr
isotopes are shown in Figs 12,13, and 14, respectively. The case of Se is
very similar to that of Ge, there are no strong deformation effects. On the
contrary, for Kr and Sr isotopes the dependence on deformation of the
uncorrelated strength distributions is huge and this is the origin of the
deformation dependence in RPA discussed earlier for these isotopes.

Figs. 15-18 show the dependence of the HF+RPA Gamow Teller strength
distributions to the Skyrme interaction used in the calculations. The
results are for SG2 (solid line) and Sk3 (dashed line) in all the isotopes
considered. In Fig. 15 for Ge isotopes we can see that there is almost no
difference in going from one interaction to another and thus, the
conclusions have a general validity. Fig. 16 for Se isotopes shows the same
characteristics. The profiles obtained with both interactions are quite
similar. The larger discrepancies occur in the prolate solutions of $^{74}$%
Se and $^{72}$Se, but this is mainly due to the different minima obtained
for these two nuclei with the two interactions (see Fig. 2), while Sk3
produces well deformed prolate solutions, SG2 has an almost spherical
solution for these two isotopes. Fig. 17 shows the results for Kr isotopes.
Here again, the strength distributions obtained with the two interactions
are quite similar in the cases where the HF solutions appear at about the
same deformation. On the other hand, when the HF solutions occurred at
different deformations in Sk3 and SG2, the strength distributions obtained
from those solutions are also quite different. This is clearly the case in $%
^{78}$Kr, where Sk3 has two solutions oblate and prolate while SG2 has a
single spherical solution. This is also true to a lesser extent in the
oblate solutions of $^{76}$Kr and $^{74}$Kr and in the prolate solution of $%
^{72}$Kr that occur at different deformations. On the other hand, the rest
of cases have very similar strength distributions and they also have very
similar deformations in the HF solutions (see Fig. 3). In Fig. 18 we can see
the results for Sr isotopes. The profiles of the strength distributions are
in this case practically the same in accordance with the situation in Fig.
4, where the HF solutions with the two interactions occur at the same
deformations.

\subsection{Half-lives and Summed Strengths}

In this subsection we present the results obtained for other quantities of
interest such as the half-lives or the GT strengths summed up to the $Q_{EC}$
window $\left( \sum_{EC}\right) $. One of the points to discuss in this
context is whether there are substantial differences in these quantities
depending on the shape of the parent nucleus. In the affirmative case, this
implies that experimental data on $\beta -$decay can be taken as a signature
of the nuclear shape. Thus, it is instructive to see the predictions for the
half-lives and summed strengths corresponding to the different stable shapes.

The total half-life $T_{1/2}$ for allowed $\beta $ decay from the ground
state of the parent nucleus is given by summing over all the final states
involved in the process

\begin{equation}
T_{1/2}^{-1}=\frac{\kappa ^{2}}{D}\sum_{\omega }f\left( Z,\omega \right)
\left| \left\langle 1_{\omega }^{+}\left\| \beta ^{+}\right\|
0^{+}\right\rangle \right| ^{2}  \label{t12}
\end{equation}
The Fermi integrals $f\left( Z,\omega \right) $ are taken from Ref. \cite
{gove}. We use $D=6200$ s and include effective factors

\begin{equation}
\kappa ^{2}=\left[ \left( g_{A}/g_{V}\right) _{eff}\right] ^{2}=\left[
0.77\left( g_{A}/g_{V}\right) _{free}\right] ^{2}=0.90  \label{quen}
\end{equation}
to take into account in an effective way the quenching of the GT coupling
constant $g_{A}$ in the nuclear medium. Note that the value of the standard
quenching factor (0.77) used here \cite{effective} is a little bit different
than the value (0.7) used in Ref. \cite{sarri98}, and consequently the
values of $T_{1/2}$ have changed in accordance.

Tables 5-7 show the results obtained from bare $2qp$, TDA, and RPA
calculations for the GT strength summed up to an energy cut of 30 MeV (Table
5), for the GT strength summed up to excitation energies below $Q_{EC}$
(Table 6), and for the total $\beta ^{+}/EC$ half-life (Table 7). The cut of
30 MeV corresponds to the excitation energy for which the Ikeda sum rule is
fulfilled up to a few per thousand. Results are shown for the two Skyrme
forces Sk3 and SG2, as well as for the different shapes oblate (o), prolate
(p), or spherical (s), where the minima occur for each isotope.

One can see in Table 5 that the summed GT strengths up to the energy cut are
conserved in going from $2qp$ to TDA calculations, but this is no longer
true in RPA where the strengths are reduced. The energy weighted sums, not
shown here, have the opposite behavior, the two quasiparticle values are
conserved by RPA while TDA produces larger energy weighted sums (see also
Ref. \cite{sarri98} ). Focussing on the RPA total strengths, one can see on
the table that the prolate shape tends to give a larger total strength, this
is more noticeable in Kr and Sr isotopes when the other equilibrium shape is
spherical. In comparing RPA results for Sk3 and SG2 one can see that the GT
summed strengths are practically equal when the predicted shapes have
similar $\beta _{0}$ values. This implies that there is a strong correlation
between the nuclear shape and the total GT strength within the $Q_{EC}$
window $\left( \sum_{EC}\right) $.

As a matter of consistency the GT excitations and $Q_{EC}$ values from which
we obtain the summed strengths $\sum_{EC}$ in Table 6 and the half-lives $%
T_{1/2\text{ }}$ in Table 7 have been calculated in the parent nucleus for
each force, shape and approach. In particular for the $Q_{EC}$ value in RPA
and TDA approximations the energy of the lowest two-quasiparticle state is
replaced by the energy $\omega $ of the lowest RPA\ or TDA state,
respectively. In most cases however the $Q_{EC}$ values in various
approaches differ at most by a few percent.

In tables 6-7, we do not include stable nuclei ($^{70}$Ge, $^{74}$Se, $^{78}$%
Kr) or nuclei near to stability with very small $Q_{EC}$ values ($^{68}$Ge, $%
^{72}$Se, $^{82}$Sr, having $Q_{EC}<0.4$ MeV, see Tables 1-4). Quantities
depending on the value of the cut $Q_{EC}$, such as those in Tables 5-6, are
extremely sensitive to the cut when $Q_{EC}$ is very small. In this case
only very few low energy excitations contribute to $\sum_{EC}$ and $T_{1/2}$
and a small change in the $Q_{EC}$ value may lead to large variations in
these quantities, specially in the half-lives that can change by orders of
magnitude. Generally, when $Q_{EC}$ is large enough small changes in $Q_{EC}$
are followed by small changes in $T_{1/2}$. This is especially true in the
deformed case, where the excitation energies are very much fragmented and
appear in an almost continuous distribution. In contrast, in the spherical
case the existence of large strengths at well located excitation energies
can make the half-lives much more dependent on fine details of the
calculations.

Note that while the half-lives in Table 7 contain already the expected
quenching factor (see Eqs. (\ref{t12},\ref{quen})), the strengths in Table 6
are in units of $\left[ g_{A}^{2}/4\pi \right] $, and therefore a reduction
of about a 50\% is expected in these strengths before comparison to
experiment is made due to the effective $g_{A}$ value.

A common feature to both tables is that the calculated half-lives $T_{1/2}$
(summed strengths $\sum_{EC}$) increase (decrease) in going from $2qp$ to
TDA to RPA. One finds variation factors of the order of ten between 2qp and
RPA calculations in $\sum_{EC}$ and $T_{1/2}$. TDA is in all cases much
closer to RPA than to 2qp calculations but still observable differences
appear in some cases. Therefore, from these calculations one concludes that
in order to achieve a reliable description of $\beta -$decay properties, an
RPA\ calculation must be performed.

Comparison to the experimental half-lives in Table 7 shows that the RPA\
results agree in general with experiment. The only exception are the $N=Z$
isotopes of Ge ($^{64}$Ge) and Se ($^{68}$Se), where we overestimate the
half-lives by a factor between 2.5 and 5. Even in the cases not included in
the table because of their small $Q_{EC}$ value ($^{68}$Ge, $^{72}$Se, $%
^{82} $Sr ), we obtain half-lives of the order of days as in experiment.

In a more detailed analysis we can see that the RPA summed strengths within
the $Q_{EC}$ window (half-lives) in the Ge isotopes are smaller (larger) for
the prolate shapes than for the oblate ones. This is due to the small peak
that appears in the distribution of the strength in the oblate cases, absent
in the prolate ones (see Figs. 7,11,15). This could lead to an observable
effect in $^{64,66}$Ge. Although the total strength (sums up to 30 MeV) are
in most cases larger for the prolate shapes, the opposite happens in the
sums cut at $Q_{EC}$. It is also important to mention that SG2 and Sk3 agree
in their predictions for the summed strengths. The results obtained for the
Se isotopes do not show any remarkable pattern and then the Se isotopes are
not good candidates to look for sizeable effects on the GT strengths due to
deformation.

Special attention deserve the cases of Kr and Sr isotopes. The summed GT
strengths up to the $Q_{EC}$ window are not conclusive to distinguish
between oblate or prolate shapes in $^{72,76}$Kr. The situation is different
in $^{74}$Kr. In this nucleus one obtains the same strength with the two
forces in the oblate case, strength which is much smaller than the strength
obtained in the calculation with the prolate shape. This fact makes $^{74}$
Kr a suitable candidate to measure its GT strength and from this measurement
to infer the ground state shape. In a similar way, $\sum_{EC}$ are about the
same in the case of $^{76}$Sr, where a coexistence between oblate and
prolate shapes is predicted. On the other hand, in the other two cases $%
^{78,80}$Sr where a prolate and spherical shape coexistence appears, $%
\sum_{EC}$ calculated in the prolate shape is clearly larger than the
corresponding strength calculated from the spherical shape. Therefore, these
two nuclei are again very interesting cases to look for these deformation
effects on the GT strengths.

\subsection{The particle-particle residual interaction}

It has often been claimed (see for instance Ref. \cite{kpp} and refs.
therein) that for a complete description of the $\beta ^{+}$ and $\beta
\beta $ strengths, the inclusion of the particle-particle (pp) residual
interaction is required. Therefore the question may arise as to why this
interaction was not included in the present work. The usual way to include
this force is in terms of a separable force with a free coupling constant $%
\kappa _{pp}$, which is fitted to the phenomenology. Since the peak of the
GTR is almost insensitive to the pp force, $\kappa _{pp}$ is usually
adjusted to reproduce the half-lives.

One of the features of the pp force is that, being an attractive force, the
GT strength is pushed down to lower energies with increasing values of $%
\kappa _{pp}$. If $\kappa _{pp}$ is strong enough it may happen that the RPA
collapses, because the condition that the ground state be stable against the
corresponding mode is not fulfilled. Inconsistencies between mean field and
residual interactions are a source of problems, particularly when discussing
single $\beta $ or double $\beta $ decays. There is work in progress to
include the pp residual force in a consistent way, starting from
Hartree-Fock-Bogoliubov calculations, where proton-neutron pairing is
included. Until this project is carried out, we have adopted in this work
the value $\kappa _{pp}=0$, which is consistent with the HF+BCS energy
density functional without proton-neutron pairing used here. Furthermore, it
has been shown \cite{hirsch} that for small values of $\kappa _{pp}$ far
from the collapse, the half-lives are nearly independent of the pp force.

Nevertheless, just as an illustration we show in this section the effect of
the inclusion of a pp force on the GT strength distributions and half-lives.
For that purpose, we introduce in our formalism a separable residual pp
force in the same way as it was done in Ref. \cite{muto92}. Using separable
forces, the QRPA equation for the separable particle-hole and
particle-particle forces can be reduced to an algebraic equation, which is
now of fourth order by the inclusion of the pp force. For the solution of
the algebraic equation we follow Ref. \cite{muto92}.

We can see in Fig. 19 the effect of the residual pp interaction on the GT
strength distributions by changing the $\kappa _{pp}$ value. The figure
corresponds to the prolate and oblate solutions of the nucleus $^{70}$Se. It
is an RPA calculation with the force SG2. As it can be seen in this figure,
the attractive character of the pp force makes the strength to be slightly
shifted to lower excitation energies, but the position of the GTR is hardly
modified by the inclusion of the pp force. We can also see in Table 8 the
total GT strength summed up to 30 MeV, as well as the sums up to $%
Q_{EC}\;\left( \sum_{EC}\right) $, and the half-lives. The total GT strength
is reduced as the value of $\kappa _{pp}$ increases, but $\sum_{EC}$
increases because of the concentration of the strength at lower energies. As
a consequence the half-lives are reduced with increasing $\kappa _{pp}$.
This is so until the collapse of the RPA takes place, which for the case
discussed here occurs at about $\kappa _{pp}=0.10$ MeV.

We also note that if we would fix $\kappa _{pp}$ to fit the experimental
value of the half-life, we would need $\kappa _{pp}=0.02$ MeV in the prolate
case and $\kappa _{pp}=0.08$ MeV in the oblate case, which is close to the
collapse. 

\section{Summary and final remarks}

We have investigated shape isomerism and $\beta -$decay in several Ge, Se,
Kr, and Sr isotopes on the basis of the selfconsistent HF+RPA framework with
Skyrme forces. This is a well founded method that has been successfully used
to describe quite diverse properties of stable spherical and deformed nuclei
through the nuclear chart. It has the appealing feature of treating the
excitations and the ground state in a selfconsistent framework with no free
parameters. This feature is particularly desirable for nuclei far from
stability, where extrapolations of methods based on local fits are more
doubtful. We took here the challenge to test the predictions of this method
on the above mentioned chains including unstable isotopes. Very reasonable
agreement with both ground state and $\beta -$decay properties is obtained.

Compared to the uncorrelated two quasiparticle response, RPA shifts the GT
strength to higher energies and reduces the total strength. While the
shifting effect is already contained in the TDA description, the quenching
effect is not. This effect produces half-lives that are much larger in RPA
than in the bare 2qp approach. Inclusion of RPA correlations are clearly
necessary for comparison to experiment.

We have found shape isomerism in most of the isotopes studied. The RPA Gamow
Teller $\beta ^{+}$ strength distributions depend on the shape (prolate,
spherical or oblate) of the parent nucleus. It is important to notice that
these results do not depend much on which effective Skyrme interaction (Sk3
or SG2) is used.

The different nuclear shapes lead in some cases to sizeable differences in
the observable range of $\beta ^{+}-$decay. We find that $^{74}$Kr, $^{78}$%
Sr and $^{80}$Sr are particularly interesting cases to look experimentally
for shape effects in $\beta ^{+}-$decay.

For even-even nuclei, neutron-proton $T=0$ and $T=1$ pairing is known to be
important when $N=Z$ \cite{nppairing}. Since our theoretical treatment does
not explicitly include neutron-proton $\left( n-p\right) $ pairing, we may
expect larger deviations between theory and experiment in the $N=Z$
isotopes. The comparison in tables 1-4 of bulk properties like binding
energies and r.m.s. radii shows that the agreement between theory and
experiment is as good for the $N=Z$ as for the $N=Z+2,Z+4,Z+6$ isotopes.
This allows us to conclude that the effect of $n-p$ pairing correlations in
the binding energy is roughly taken into account by the use of the
phenomenological gap parameters $\Delta _{p},\Delta _{n}.$ This could be
expected from HFB theory where the total gap satisfies

\[
\left| \Delta _{p}\right| ^{2}=\left| \Delta _{pp}\right| ^{2}+\left| \Delta
_{pn}^{T=1}\right| ^{2}+\left| \Delta _{pn}^{T=0}\right| ^{2} 
\]
and similarly for neutrons.

$\beta ^{+}-$decay strength functions and half-lives of $N=Z$ nuclei are
expected to be more sensitive to the explicit inclusion of $n-p$ pairing in
the microscopic calculations. Indeed, a look at our RPA results in table 7
shows that for the $N=Z$ isotopes of Ge and Se, the experimental half-lives
are overestimated by a factor of 3 to 5, depending on the interaction and
shape, while fair agreement with experimental half-lives is obtained for the 
$N>Z$ isotopes. Interestingly enough our RPA results for the $N=Z$ isotopes
of Kr and Sr are in good agreement with experiment. It will therefore be
interesting to see how the inclusion of $n-p$ pairing in our microscopic
calculation affects our present results. It will also be interesting to
compare our results with future data on $\beta ^{+}-$strengths.

\acknowledgments 

We are thankful to J. Dukelsky, M.J. Garc\'{\i }a Borge, W. Gelletly, and
Ch. Mieh\'{e} for stimulating comments and discussions. This work was
supported by DGICYT (Spain) under contract number PB95/0123. One of us
(A.E.) thanks Ministerio de Educaci\'{o}n y Cultura (Spain) for support.

\newpage

\begin{center}
{\bf Figure captions}
\end{center}

{\bf Figure 1. }Total energy of the Ge isotopes $^{64,66,68,70}$Ge as a
function of the mass quadrupole moment $Q_{0}$. The results correspond to a
constrained HF+BCS calculation with the Skyrme interaction SG2 (solid line)
and Sk3 (dashed line). The distance between two ticks in the vertical axes
is always 1 MeV but the origin is different for each curve.

{\bf Figure 2. }Same as in Fig. 1 for the Se isotopes $^{68,70,72,74}$Se.

{\bf Figure 3. }Same as in Fig. 1 for the Kr isotopes $^{72,74,76,78}$Kr.

{\bf Figure 4. }Same as in Fig. 1 for the Sr isotopes $^{76,78,80,82}$Sr.

{\bf Figure 5. }$\left( p,n\right) $ and $\left( n,p\right) $ $L=0$ cross
sections in $^{54,56}$Fe compared to theoretical GT strength distributions
obtained with the force SG2 in RPA. Experimental data for 
$\left( p,n\right) $ and $\left( n,p\right) $ reactions are from 
\cite{rapaport} and \cite{elka}, respectively.

{\bf Figure 6. }$\left( n,p\right) $ $L=0$ cross sections in $^{70,72}$Ge 
\cite{vetterli} compared with the RPA theoretical GT strength
distributions obtained from SG2.

{\bf Figure 7. }Comparison of the Gamow Teller strength distribution 
[$g_{A}^{2}/4\pi $] in the Ge isotopes $^{64,66,68,70}$Ge. The results are for
the force SG2 in RPA.

{\bf Figure 8. }Same as in Fig. 7 for the Se isotopes $^{68,70,72,74}$Se.

{\bf Figure 9. }Same as in Fig. 7 for the Kr isotopes $^{72,74,76,78}$Kr.

{\bf Figure 10. }Same as in Fig. 7 for the Sr isotopes $^{76,78,80,82}$Sr.

{\bf Figure 11. }Comparison of RPA (solid line), TDA (dashed line), and bare
two-quasiparticle (dotted line) Gamow Teller strength distributions 
[$g_{A}^{2}/4\pi $] in the Ge isotopes $^{64,66,68,70}$Ge. The results
correspond to the force SG2.

{\bf Figure 12. }Same as in Fig. 11 for the Se isotopes $^{68,70,72,74}$Se.

{\bf Figure 13. }Same as in Fig. 11 for the Kr isotopes $^{72,74,76,78}$Kr.

{\bf Figure 14. }Same as in Fig. 11 for the Sr isotopes $^{76,78,80,82}$Sr.

{\bf Figure 15. }Gamow Teller strength distributions [$g_{A}^{2}/4\pi $] in
the Ge isotopes $^{64,66,68,70}$Ge as a function of the excitation energy of
the daughter nucleus. The results correspond to the forces SG2 (solid line)
and Sk3 (dashed line) in RPA.

{\bf Figure 16. }Same as in Fig. 15 for the Se isotopes $^{68,70,72,74}$Se.

{\bf Figure 17. }Same as in Fig. 15 for the Kr isotopes $^{72,74,76,78}$Kr.

{\bf Figure 18. }Same as in Fig. 15 for the Sr isotopes $^{76,78,80,82}$Sr.

{\bf Figure 19. } GT strength distributions in $^{70}$Se calculated in RPA
with the force SG2 for various values of the coupling strength $\kappa _{pp}$
of the particle-particle force.

\newpage

\begin{table}[tbp]
{\bf Table 1. }Pairing gap parameters for neutrons and protons $\Delta
_{n},\Delta _{p}$ [MeV]; Fermi energies for neutrons and protons $\lambda
_{n},\lambda _{p}$ [MeV]; charge radii $r_{C}$ [fm]; charge $Q_{0,p}$ and
mass $Q_{0}$ quadrupole moments [fm$^{2}$]; values of the deformation
parameters $\beta _{0}$, $\left\langle J^{2}\right\rangle $, cranking
moments of inertia ${\cal I}_{cr}$ [MeV$^{-1}$], gyromagnetic ratios $g_{R}$
, binding energies $E_{T}$ [MeV], coupling strengths of the residual
spin-isospin interaction $\chi _{GT}$ [MeV], and $Q_{EC}$ [MeV] values for
the chain of Ge isotopes $^{64,66,68,70}$Ge. Experimental values for $%
r_{C},\beta _{0},E_{T},Q_{EC\text{ }}$are from Ref. \cite{deVries,raman,audi}%
, respectively.
\par
\begin{tabular}{cccccccccrccccccccccc}
&  &  & $\Delta _{n}$ & $\Delta _{p}$ & $\lambda _{n}$ & $\lambda _{p}$ & 
\multicolumn{2}{c}{$r_{C}$} & $Q_{0,p}$ & $Q_{0}$ & \multicolumn{2}{c}{$%
\beta _{0}$} & $\left\langle J^{2}\right\rangle $ & ${\cal I}_{cr}$ & $g_{R}$
& \multicolumn{2}{c}{$E_{T}$} & $\chi _{GT}$ & \multicolumn{2}{c}{$Q_{EC}$}
\\ 
\multicolumn{1}{l}{} & \multicolumn{1}{l}{} & \multicolumn{1}{l}{} & 
\multicolumn{1}{l}{} & \multicolumn{1}{l}{} & \multicolumn{1}{l}{} & 
\multicolumn{1}{l}{} & th. & exp. & \multicolumn{1}{l}{} & 
\multicolumn{1}{l}{} & th. & exp. & \multicolumn{1}{l}{} & 
\multicolumn{1}{l}{} & \multicolumn{1}{l}{} & th. & exp. &  & th. & exp. \\ 
\hline
&  &  & 2.10 & 1.80 &  &  &  &  &  &  &  &  & \multicolumn{1}{r}{} &  &  & 
& -545.9 &  &  & 4.41 \\ 
$^{64}$Ge & oblate & SG2 &  &  & -13.43 & -3.42 & 4.01 & \multicolumn{1}{r}{}
& -122 & \multicolumn{1}{r}{-238} & \multicolumn{1}{r}{-0.19} &  & 
\multicolumn{1}{r}{14.8} & 2.0 & 0.57 & -558.4 &  & 0.43 & 
\multicolumn{1}{r}{4.3} &  \\ 
&  & Sk3 &  &  & -13.09 & -3.09 & 4.03 & \multicolumn{1}{r}{} & -111 & 
\multicolumn{1}{r}{-217} & \multicolumn{1}{r}{-0.17} &  & \multicolumn{1}{r}{
12.4} & 1.7 & 0.56 & -541.6 &  & 0.42 & \multicolumn{1}{r}{4.2} &  \\ 
& prolate & SG2 &  &  & -13.32 & -3.28 & 4.00 & \multicolumn{1}{r}{} & 121 & 
\multicolumn{1}{r}{236} & \multicolumn{1}{r}{0.19} &  & \multicolumn{1}{r}{
12.4} & 1.8 & 0.57 & -558.4 &  & 0.43 & \multicolumn{1}{r}{4.1} &  \\ 
&  & Sk3 &  &  & -13.05 & -3.02 & 4.03 & \multicolumn{1}{r}{} & 128 & 
\multicolumn{1}{r}{249} & \multicolumn{1}{r}{0.20} &  & \multicolumn{1}{r}{
13.5} & 1.9 & 0.56 & -542.2 &  & 0.42 & \multicolumn{1}{r}{3.9} &  \\ \hline
&  &  & 1.80 & 1.60 &  &  &  &  &  &  &  & $\pm 0.17$ & \multicolumn{1}{r}{}
&  &  &  & -569.3 &  & \multicolumn{1}{r}{} & 2.10 \\ 
$^{66}$Ge & oblate & SG2 &  &  & -12.35 & -4.60 & 4.02 & \multicolumn{1}{r}{}
& -134 & \multicolumn{1}{r}{-272} & \multicolumn{1}{r}{-0.21} &  & 
\multicolumn{1}{r}{20.2} & 3.1 & 0.50 & -582.3 &  & 0.41 & 
\multicolumn{1}{r}{2.1} &  \\ 
&  & Sk3 &  &  & -11.99 & -4.48 & 4.06 & \multicolumn{1}{r}{} & -132 & 
\multicolumn{1}{r}{-270} & \multicolumn{1}{r}{-0.21} &  & \multicolumn{1}{r}{
19.6} & 2.9 & 0.47 & -564.8 &  & 0.40 & \multicolumn{1}{r}{1.9} &  \\ 
& prolate & SG2 &  &  & -12.21 & -4.40 & 4.01 & \multicolumn{1}{r}{} & 125 & 
\multicolumn{1}{r}{251} & \multicolumn{1}{r}{0.20} &  & \multicolumn{1}{r}{
15.1} & 2.5 & 0.52 & -581.8 &  & 0.41 & \multicolumn{1}{r}{2.1} &  \\ 
&  & Sk3 &  &  & -11.83 & -4.32 & 4.05 & \multicolumn{1}{r}{} & 135 & 
\multicolumn{1}{r}{274} & \multicolumn{1}{r}{0.21} &  & \multicolumn{1}{r}{
169.} & 2.6 & 0.51 & -564.8 &  & 0.40 & \multicolumn{1}{r}{1.7} &  \\ \hline
&  &  & 1.90 & 1.60 &  &  &  &  &  &  &  & $\pm 0.20$ & \multicolumn{1}{r}{}
&  &  &  & -590.8 &  & \multicolumn{1}{r}{} & 0.11 \\ 
$^{68}$Ge & oblate & SG2 &  &  & -11.39 & -5.69 & 4.04 & \multicolumn{1}{r}{}
& -131 & \multicolumn{1}{r}{-274} & \multicolumn{1}{r}{-0.20} &  & 
\multicolumn{1}{r}{18.8} & 2.8 & 0.53 & -604.8 &  & 0.40 & 
\multicolumn{1}{r}{0.1} &  \\ 
&  & Sk3 &  &  & -11.02 & -5.76 & 4.08 & \multicolumn{1}{r}{} & -135 & 
\multicolumn{1}{r}{-285} & \multicolumn{1}{r}{-0.21} &  & \multicolumn{1}{r}{
20.0} & 2.9 & 0.46 & -586.7 &  & 0.39 & \multicolumn{1}{r}{-0.5} &  \\ 
& prolate & SG2 &  &  & -11.36 & -5.48 & 4.02 & \multicolumn{1}{r}{} & 116 & 
\multicolumn{1}{r}{237} & \multicolumn{1}{r}{0.18} &  & \multicolumn{1}{r}{
13.0} & 2.1 & 0.56 & -604.1 &  & 0.40 & \multicolumn{1}{r}{0.3} &  \\ 
&  & Sk3 &  &  & -10.89 & -5.57 & 4.07 & \multicolumn{1}{r}{} & 136 & 
\multicolumn{1}{r}{285} & \multicolumn{1}{r}{0.21} &  & \multicolumn{1}{r}{
17.2} & 2.6 & 0.50 & -586.4 &  & 0.39 & \multicolumn{1}{r}{-0.6} &  \\ \hline
&  &  & 1.90 & 1.60 &  &  &  & 4.05 &  &  &  & $\pm 0.22$ & 
\multicolumn{1}{r}{} &  &  &  & -610.5 &  & \multicolumn{1}{r}{} & -- \\ 
$^{70}$Ge & oblate & SG2 &  &  & -10.70 & -6.65 & 4.04 & \multicolumn{1}{r}{}
& -111 & \multicolumn{1}{r}{-236} & \multicolumn{1}{r}{-0.17} &  & 
\multicolumn{1}{r}{13.6} & 2.2 & 0.55 & -625.6 &  & 0.39 & 
\multicolumn{1}{r}{-1.3} &  \\ 
&  & Sk3 &  &  & -10.27 & -6.84 & 4.09 & \multicolumn{1}{r}{} & -118 & 
\multicolumn{1}{r}{-255} & \multicolumn{1}{r}{-0.18} &  & \multicolumn{1}{r}{
15.4} & 2.3 & 0.48 & -606.8 &  & 0.38 & \multicolumn{1}{r}{-2.1} &  \\ 
& prolate & SG2 &  &  & -10.83 & -6.44 & 4.03 & \multicolumn{1}{r}{} & 78 & 
\multicolumn{1}{r}{157} & \multicolumn{1}{r}{0.11} &  & \multicolumn{1}{r}{
5.6} & 1.0 & 0.68 & -624.8 &  & 0.39 & \multicolumn{1}{r}{-1.1} &  \\ 
&  & Sk3 &  &  & -10.18 & -6.77 & 4.09 & \multicolumn{1}{r}{} & 144 & 
\multicolumn{1}{r}{315} & \multicolumn{1}{r}{0.22} &  & \multicolumn{1}{r}{
18.8} & 2.8 & 0.46 & -606.1 &  & 0.38 & \multicolumn{1}{r}{-2.0} & 
\end{tabular}
\end{table}

\newpage

\begin{table}[tbp]
{\bf Table 2. }Same as in Table 1 for the Se isotopes $^{68,70,72,74}$Se.
\par
\begin{tabular}{ccccccccrcllcccllcll}
&  &  & $\Delta _{n}$ & $\Delta _{p}$ & $\lambda _{n}$ & $\lambda _{p}$ & $%
r_{C}$ & $Q_{0,p}$ & $Q_{0}$ & \multicolumn{2}{c}{$\beta _{0}$} & $%
\left\langle J^{2}\right\rangle $ & ${\cal I}_{cr}$ & $g_{R}$ & 
\multicolumn{2}{c}{$E_{T}$} & $\chi _{GT}$ & \multicolumn{2}{c}{$Q_{EC}$} \\ 
\multicolumn{1}{l}{} & \multicolumn{1}{l}{} & \multicolumn{1}{l}{} & 
\multicolumn{1}{l}{} & \multicolumn{1}{l}{} & \multicolumn{1}{l}{} & 
\multicolumn{1}{l}{} & \multicolumn{1}{l}{} & \multicolumn{1}{l}{} & 
\multicolumn{1}{l}{} & th. & exp. & \multicolumn{1}{l}{} & 
\multicolumn{1}{l}{} & \multicolumn{1}{l}{} & th. & exp. & 
\multicolumn{1}{l}{} & th. & exp. \\ \hline
$^{68}$Se &  &  & 2.20 & 1.80 &  &  &  &  &  & \multicolumn{1}{c}{} & 
\multicolumn{1}{c}{} & \multicolumn{1}{r}{} &  &  & \multicolumn{1}{c}{} & 
\multicolumn{1}{c}{-576.4} &  & \multicolumn{1}{c}{} & \multicolumn{1}{c}{
4.70} \\ 
& oblate & SG2 &  &  & -13.61 & -3.15 & 4.10 & -153 & \multicolumn{1}{r}{-299
} & \multicolumn{1}{r}{-0.22} & \multicolumn{1}{c}{} & \multicolumn{1}{r}{
21.1} & 3.0 & 0.57 & \multicolumn{1}{c}{-590.3} & \multicolumn{1}{c}{} & 0.40
& \multicolumn{1}{r}{4.1} & \multicolumn{1}{c}{} \\ 
&  & Sk3 &  &  & -13.43 & -2.98 & 4.13 & -154 & \multicolumn{1}{r}{-302} & 
\multicolumn{1}{r}{-0.22} & \multicolumn{1}{c}{} & \multicolumn{1}{r}{21.3}
& 3.0 & 0.57 & \multicolumn{1}{c}{-572.6} & \multicolumn{1}{c}{} & 0.39 & 
\multicolumn{1}{r}{4.2} & \multicolumn{1}{c}{} \\ 
& prolate & SG2 &  &  & -13.49 & -2.98 & 4.08 & 134 & \multicolumn{1}{r}{262}
& \multicolumn{1}{r}{0.19} & \multicolumn{1}{c}{} & \multicolumn{1}{r}{14.9}
& 2.2 & 0.59 & \multicolumn{1}{c}{-589.2} & \multicolumn{1}{c}{} & 0.40 & 
\multicolumn{1}{r}{4.4} & \multicolumn{1}{c}{} \\ 
&  & Sk3 &  &  & -13.30 & -2.82 & 4.12 & 145 & \multicolumn{1}{r}{283} & 
\multicolumn{1}{r}{0.20} & \multicolumn{1}{c}{} & \multicolumn{1}{r}{16.6} & 
2.3 & 0.58 & \multicolumn{1}{c}{-572.0} & \multicolumn{1}{c}{} & 0.39 & 
\multicolumn{1}{r}{4.2} & \multicolumn{1}{c}{} \\ \hline
$^{70}$Se &  &  & 1.80 & 1.80 &  &  &  &  &  & \multicolumn{1}{c}{} & 
\multicolumn{1}{c}{$\pm 0.29$} & \multicolumn{1}{r}{} &  &  & 
\multicolumn{1}{c}{} & \multicolumn{1}{c}{-600.3} &  & \multicolumn{1}{r}{}
& \multicolumn{1}{c}{2.40} \\ 
& oblate & SG2 &  &  & -12.36 & -4.26 & 4.11 & -155 & \multicolumn{1}{r}{-311
} & \multicolumn{1}{r}{-0.22} & \multicolumn{1}{c}{} & \multicolumn{1}{r}{
22.9} & 3.4 & 0.50 & \multicolumn{1}{c}{-614.1} & \multicolumn{1}{c}{} & 0.39
& \multicolumn{1}{r}{2.2} & \multicolumn{1}{c}{} \\ 
&  & Sk3 &  &  & -12.14 & -4.24 & 4.15 & -161 & \multicolumn{1}{r}{-328} & 
\multicolumn{1}{r}{-0.23} & \multicolumn{1}{c}{} & \multicolumn{1}{r}{25.0}
& 3.7 & 0.46 & \multicolumn{1}{c}{-596.1} & \multicolumn{1}{c}{} & 0.38 & 
\multicolumn{1}{r}{2.1} & \multicolumn{1}{c}{} \\ 
& prolate & SG2 &  &  & -12.36 & -4.11 & 4.09 & 126 & \multicolumn{1}{r}{251}
& \multicolumn{1}{r}{0.18} & \multicolumn{1}{c}{} & \multicolumn{1}{r}{14.5}
& 2.3 & 0.52 & \multicolumn{1}{c}{-612.9} & \multicolumn{1}{c}{} & 0.39 & 
\multicolumn{1}{r}{2.6} & \multicolumn{1}{c}{} \\ 
&  & Sk3 &  &  & -12.04 & -4.11 & 4.14 & 153 & \multicolumn{1}{r}{310} & 
\multicolumn{1}{r}{0.21} & \multicolumn{1}{c}{} & \multicolumn{1}{r}{20.1} & 
2.9 & 0.48 & \multicolumn{1}{c}{-595.1} & \multicolumn{1}{c}{} & 0.38 & 
\multicolumn{1}{r}{2.0} & \multicolumn{1}{c}{} \\ \hline
$^{72}$Se &  &  & 1.80 & 1.80 &  &  &  &  &  & \multicolumn{1}{c}{} & 
\multicolumn{1}{c}{$\pm 0.21$} & \multicolumn{1}{r}{} &  &  & 
\multicolumn{1}{c}{} & \multicolumn{1}{c}{-622.4} &  & \multicolumn{1}{r}{}
& \multicolumn{1}{c}{0.34} \\ 
& oblate & SG2 &  &  & -11.68 & -5.28 & 4.11 & -129 & \multicolumn{1}{r}{-264
} & \multicolumn{1}{r}{-0.18} & \multicolumn{1}{c}{} & \multicolumn{1}{r}{
16.2} & 2.6 & 0.51 & \multicolumn{1}{c}{-636.8} & \multicolumn{1}{c}{} & 0.38
& \multicolumn{1}{r}{0.7} & \multicolumn{1}{c}{} \\ 
&  & Sk3 &  &  & -11.36 & -5.43 & 4.16 & -156 & \multicolumn{1}{r}{-326} & 
\multicolumn{1}{r}{-0.21} & \multicolumn{1}{c}{} & \multicolumn{1}{r}{23.1}
& 3.4 & 0.45 & \multicolumn{1}{c}{-618.3} & \multicolumn{1}{c}{} & 0.37 & 
\multicolumn{1}{r}{0.3} & \multicolumn{1}{c}{} \\ 
& prolate & SG2 &  &  & -11.89 & -5.22 & 4.09 & 76 & \multicolumn{1}{r}{148}
& \multicolumn{1}{r}{0.10} & \multicolumn{1}{c}{} & \multicolumn{1}{r}{4.9}
& 0.9 & 0.63 & \multicolumn{1}{c}{-635.6} & \multicolumn{1}{c}{} & 0.38 & 
\multicolumn{1}{r}{1.2} & \multicolumn{1}{c}{} \\ 
&  & Sk3 &  &  & -11.44 & -5.56 & 4.20 & 248 & \multicolumn{1}{r}{526} & 
\multicolumn{1}{r}{0.34} & \multicolumn{1}{c}{} & \multicolumn{1}{r}{42.0} & 
5.7 & 0.38 & \multicolumn{1}{c}{-617.5} & \multicolumn{1}{c}{} & 0.37 & 
\multicolumn{1}{r}{0.4} & \multicolumn{1}{c}{} \\ \hline
$^{74}$Se &  &  & 1.80 & 1.80 &  &  &  &  &  & \multicolumn{1}{c}{} & 
\multicolumn{1}{c}{$\pm 0.30$} & \multicolumn{1}{r}{} &  &  & 
\multicolumn{1}{c}{} & \multicolumn{1}{c}{-642.9} &  & \multicolumn{1}{r}{}
& \multicolumn{1}{c}{-} \\ 
& oblate & SG2 &  &  & -10.97 & -6.33 & 4.13 & -131 & \multicolumn{1}{r}{-277
} & \multicolumn{1}{r}{-0.18} & \multicolumn{1}{c}{} & \multicolumn{1}{r}{
16.6} & 2.7 & 0.45 & \multicolumn{1}{c}{-658.6} & \multicolumn{1}{c}{} & 0.37
& \multicolumn{1}{r}{-1.2} & \multicolumn{1}{c}{} \\ 
&  & Sk3 &  &  & -10.66 & -6.52 & 4.17 & -134 & \multicolumn{1}{r}{-285} & 
\multicolumn{1}{r}{-0.18} & \multicolumn{1}{c}{} & \multicolumn{1}{r}{17.5}
& 2.8 & 0.42 & \multicolumn{1}{c}{-639.5} & \multicolumn{1}{c}{} & 0.36 & 
\multicolumn{1}{r}{-2.0} & \multicolumn{1}{c}{} \\ 
& spherical & SG2 &  &  & -11.26 & -6.29 & 4.10 & \multicolumn{1}{c}{0} & 0
& \multicolumn{1}{c}{0} & \multicolumn{1}{c}{} & 0 & 0 & -- & 
\multicolumn{1}{c}{-658.2} & \multicolumn{1}{c}{} & 0.37 & 
\multicolumn{1}{r}{-1.0} & \multicolumn{1}{c}{} \\ 
& prolate & Sk3 &  &  & -10.55 & -6.57 & 4.20 & 226 & \multicolumn{1}{r}{493}
& \multicolumn{1}{r}{0.31} & \multicolumn{1}{c}{} & 36.3 & 5.0 & 0.37 & 
\multicolumn{1}{c}{-638.4} & \multicolumn{1}{c}{} & 0.36 & 
\multicolumn{1}{r}{-2.0} & \multicolumn{1}{c}{}
\end{tabular}
\end{table}
\newpage

\begin{table}[tbp]
{\bf Table 3. }Same as in Table 1 for the Kr isotopes $^{72,74,76,78}$Kr.
Experimental values for $r_{C}$ are from Ref. \cite{keim}.
\par
\begin{tabular}{cccccccccrccccccccccc}
&  &  & $\Delta _{n}$ & $\Delta _{p}$ & $\lambda _{n}$ & $\lambda _{p}$ & 
\multicolumn{2}{c}{$r_{C}$} & $Q_{0,p}$ & $Q_{0}$ & \multicolumn{2}{c}{$%
\beta _{0}$} & $\left\langle J^{2}\right\rangle $ & ${\cal I}_{cr}$ & $g_{R}$
& \multicolumn{2}{c}{$E_{T}$} & $\chi _{GT}$ & \multicolumn{2}{c}{$Q_{EC}$}
\\ 
&  &  &  &  &  &  & th. & exp. &  & \multicolumn{1}{r}{} & th. & exp. &  & 
&  & th. & exp. &  & th. & exp. \\ \hline
$^{72}$Kr &  &  & 1.50 & 1.50 &  &  &  & 4.163 &  &  &  &  &  &  &  &  & 
-607.1 &  &  & 5.04 \\ 
& oblate & SG2 &  &  & -13.24 & -2.57 & 4.17 &  & -189 & \multicolumn{1}{r}{
-367} & \multicolumn{1}{r}{-0.25} &  & 31.8 & 5.1 & 0.52 & -618.5 &  & 0.38
& 4.9 &  \\ 
&  & Sk3 &  &  & -13.26 & -2.64 & 4.22 &  & -210 & \multicolumn{1}{r}{-412}
& \multicolumn{1}{r}{-0.27} &  & 39.3 & 6.1 & 0.52 & -601.2 &  & 0.37 & 4.9
&  \\ 
& prolate & SG2 &  &  & -13.44 & -2.65 & 4.13 &  & 108 & \multicolumn{1}{r}{
210} & \multicolumn{1}{r}{0.14} &  & 11.9 & 2.3 & 0.51 & -617.2 &  & 0.38 & 
5.2 &  \\ 
&  & Sk3 &  &  & -13.36 & -2.73 & 4.22 &  & 237 & \multicolumn{1}{r}{465} & 
\multicolumn{1}{r}{0.30} &  & 38.6 & 5.8 & 0.51 & -599.9 &  & 0.37 & 5.2 & 
\\ \hline
$^{74}$Kr &  &  & 1.50 & 1.50 &  &  &  & 4.187 &  &  &  & $\pm 0.39$ &  &  & 
&  & -631.3 &  &  & 3.14 \\ 
& oblate & SG2 &  &  & -12.74 & -3.90 & 4.15 &  & -112 & \multicolumn{1}{r}{
-222} & \multicolumn{1}{r}{-0.15} &  & 13.4 & 2.4 & 0.50 & -643.9 &  & 0.37
& 3.2 &  \\ 
&  & Sk3 &  &  & -12.30 & -3.80 & 4.23 &  & -204 & \multicolumn{1}{r}{-412}
& \multicolumn{1}{r}{-0.26} &  & 21.5 & 3.6 & 0.49 & -625.4 &  & 0.36 & 3.0
&  \\ 
& prolate & SG2 &  &  & -12.75 & -4.14 & 4.23 &  & 307 & \multicolumn{1}{r}{
623} & \multicolumn{1}{r}{0.39} &  & 60.2 & 9.1 & 0.46 & -642.9 &  & 0.37 & 
3.3 &  \\ 
&  & Sk3 &  &  & -12.59 & -4.29 & 4.27 &  & 311 & \multicolumn{1}{r}{633} & 
\multicolumn{1}{r}{0.39} &  & 60.3 & 8.9 & 0.46 & -625.9 &  & 0.36 & 2.7 & 
\\ \hline
$^{76}$Kr &  &  & 1.60 & 1.70 &  &  &  & 4.202 &  & \multicolumn{1}{r}{} & 
\multicolumn{1}{r}{} & $\pm 0.41$ &  &  &  &  & -654.2 &  &  & 1.31 \\ 
& spherical & SG2 &  &  & -12.23 & -5.18 & 4.15 &  & \multicolumn{1}{c}{0} & 
0 & 0 &  & 0 & 0 & -- & -668.5 &  & 0.36 & 1.4 &  \\ 
& oblate & Sk3 &  &  & -11.76 & -5.01 & 4.21 &  & -142 & \multicolumn{1}{r}{
-294} & \multicolumn{1}{r}{-0.18} &  & 18.4 & 3.0 & 0.40 & -649.3 &  & 0.35
& 0.9 &  \\ 
& prolate & SG2 &  &  & -11.76 & -5.09 & 4.25 &  & 299 & \multicolumn{1}{r}{
623} & \multicolumn{1}{r}{0.37} &  & 54.9 & 4.9 & 0.44 & -666.7 &  & 0.36 & 
1.1 &  \\ 
&  & Sk3 &  &  & -11.36 & -5.37 & 4.29 &  & 302 & \multicolumn{1}{r}{634} & 
\multicolumn{1}{r}{0.37} &  & 56.3 & 7.7 & 0.44 & -649.2 &  & 0.35 & 0.6 & 
\\ \hline
$^{78}$Kr &  &  & 1.70 & 1.80 &  &  &  & 4.204 &  &  &  & $\pm 0.34$ &  &  & 
&  & -675.6 &  &  & -- \\ 
& spherical & SG2 &  &  & -11.73 & -6.18 & 4.16 &  & \multicolumn{1}{c}{0} & 
0 & 0 &  & 0 & 0 & -- & -691.3 &  & 0.35 & -0.7 &  \\ 
& oblate & Sk3 &  &  & -10.99 & -6.17 & 4.23 &  & -142 & \multicolumn{1}{r}{
-301} & \multicolumn{1}{r}{-0.18} &  & 17.8 & 2.9 & 0.38 & -671.7 &  & 0.34
& -0.9 &  \\ 
& prolate & Sk3 &  &  & -10.81 & -6.22 & 4.28 &  & 256 & \multicolumn{1}{r}{
546} & \multicolumn{1}{r}{0.31} &  & 41.1 & 5.7 & 0.44 & -670.6 &  & 0.34 & 
-1.1 & 
\end{tabular}
\end{table}
\newpage

\begin{table}[tbp]
{\bf Table 4. }Same as in Table 1 for the Sr isotopes $^{76,78,80,82}$Sr.
Experimental values for $r_{C}$ are from Ref. \cite{buchinger}.
\par
\begin{tabular}{cccccccccrccccccccccc}
&  &  & $\Delta _{n}$ & $\Delta _{p}$ & $\lambda _{n}$ & $\lambda _{p}$ & 
\multicolumn{2}{c}{$r_{C}$} & $Q_{0,p}$ & $Q_{0}$ & \multicolumn{2}{c}{$%
\beta _{0}$} & $\left\langle J^{2}\right\rangle $ & ${\cal I}_{cr}$ & $g_{R}$
& \multicolumn{2}{c}{$E_{T}$} & $\chi _{GT}$ & \multicolumn{2}{c}{$Q_{EC}$}
\\ 
&  &  &  &  &  &  & th. & exp. &  & \multicolumn{1}{r}{} & th. & exp. &  & 
&  & th. & exp. &  & th. & exp. \\ \hline
&  &  & 1.50 & 1.50 &  &  &  &  &  &  &  &  &  & \multicolumn{1}{r}{} &  & 
& -638.1 &  & \multicolumn{1}{r}{} & 6.10 \\ 
$^{76}$Sr & oblate & SG2 &  &  & -13.87 & -2.66 & 4.20 &  & -94 & 
\multicolumn{1}{r}{-183} & \multicolumn{1}{r}{-0.11} &  & \multicolumn{1}{r}{
9.4} & \multicolumn{1}{r}{1.9} & 0.51 & -649.9 &  & 0.36 & 
\multicolumn{1}{r}{5.2} &  \\ 
&  & Sk3 &  &  & -13.77 & -2.59 & 4.23 &  & -109 & \multicolumn{1}{r}{-213}
& \multicolumn{1}{r}{-0.13} &  & \multicolumn{1}{r}{11.3} & 
\multicolumn{1}{r}{2.0} & 0.51 & -630.6 &  & 0.35 & \multicolumn{1}{r}{5.7}
&  \\ 
& prolate & SG2 &  &  & -13.78 & -2.67 & 4.30 &  & 359 & \multicolumn{1}{r}{
702} & \multicolumn{1}{r}{0.42} &  & \multicolumn{1}{r}{69.7} & 
\multicolumn{1}{r}{10.3} & 0.51 & -649.5 &  & 0.36 & \multicolumn{1}{r}{5.2}
&  \\ 
&  & Sk3 &  &  & -13.77 & -2.71 & 4.34 &  & 358 & \multicolumn{1}{r}{703} & 
\multicolumn{1}{r}{0.41} &  & \multicolumn{1}{r}{69.5} & \multicolumn{1}{r}{
10.0} & 0.51 & -632.8 &  & 0.35 & \multicolumn{1}{r}{4.9} &  \\ \hline
&  &  & 1.30 & 1.30 &  &  &  & 4.217 &  &  &  & $\pm 0.43$ &  &  &  &  & 
-663.0 &  & \multicolumn{1}{r}{} & 3.76 \\ 
$^{78}$Sr & spherical & SG2 &  &  & -13.19 & -4.02 & 4.19 &  & 
\multicolumn{1}{c}{0} & 0 & 0 &  & 0 & 0 & -- & -676.2 &  & 0.35 & 
\multicolumn{1}{r}{3.9} &  \\ 
&  & Sk3 &  &  & -12.98 & -3.90 & 4.23 &  & \multicolumn{1}{c}{0} & 0 & 0 & 
& 0 & 0 & -- & -656.0 &  & 0.34 & \multicolumn{1}{r}{3.2} &  \\ 
& prolate & SG2 &  &  & -12.46 & -3.62 & 4.31 &  & 364 & \multicolumn{1}{r}{
732} & \multicolumn{1}{r}{0.42} &  & 74.5 & 11.3 & 0.54 & -674.3 &  & 0.35 & 
\multicolumn{1}{r}{2.7} &  \\ 
&  & Sk3 &  &  & -12.25 & -3.83 & 4.35 &  & 369 & \multicolumn{1}{r}{746} & 
\multicolumn{1}{r}{0.42} &  & 75.7 & 11.0 & 0.54 & -657.5 &  & 0.34 & 
\multicolumn{1}{r}{3.7} &  \\ \hline
&  &  & 1.60 & 1.60 &  &  &  & 4.217 &  &  &  & $\pm 0.38$ &  &  &  &  & 
-686.3 &  & \multicolumn{1}{r}{} & 1.87 \\ 
$^{80}$Sr & spherical & SG2 &  &  & -12.19 & -4.92 & 4.21 &  & 
\multicolumn{1}{c}{0} & 0 & 0 &  & 0 & 0 & -- & -701.4 &  & 0.34 & 
\multicolumn{1}{r}{1.5} &  \\ 
&  & Sk3 &  &  & -12.22 & -4.99 & 4.25 &  & \multicolumn{1}{c}{0} & 0 & 0 & 
& 0 & 0 & -- & -681.4 &  & 0.33 & \multicolumn{1}{r}{0.8} &  \\ 
& prolate & SG2 &  &  & -11.88 & -4.51 & 4.31 &  & 336 & \multicolumn{1}{r}{
686} & \multicolumn{1}{r}{0.38} &  & 60.4 & 8.7 & 0.53 & -698.1 &  & 0.34 & 
\multicolumn{1}{r}{1.4} &  \\ 
&  & Sk3 &  &  & -11.46 & -4.77 & 4.36 &  & 340 & \multicolumn{1}{r}{699} & 
\multicolumn{1}{r}{0.39} &  & 61.0 & 8.5 & 0.53 & -680.6 &  & 0.33 & 
\multicolumn{1}{r}{1.7} &  \\ \hline
&  &  & 1.70 & 1.80 &  &  &  & 4.209 &  &  &  & $\pm 0.29$ &  &  &  &  & 
-708.1 &  & \multicolumn{1}{r}{} & 0.18 \\ 
$^{82}$Sr & spherical & SG2 &  &  & -11.60 & -5.86 & 4.23 &  & 
\multicolumn{1}{c}{0} & 0 & 0 &  & 0 & 0 & -- & -724.8 &  & 0.33 & 
\multicolumn{1}{r}{0.1} &  \\ 
&  & Sk3 &  &  & -11.57 & -6.03 & 4.27 &  & \multicolumn{1}{c}{0} & 0 & 0 & 
& 0 & 0 & -- & -705.2 &  & 0.32 & \multicolumn{1}{r}{-0.3} & 
\end{tabular}
\end{table}

\newpage

\begin{table}[tbp]
{\bf Table 5. }Results from bare $2qp$, TDA, and RPA calculations for the
Gamow Teller strength in units of $\left[ g_{A}^{2}/4\pi \right] $ summed up
to $E_{cut}=30$ MeV. The results correspond to the two Skyrme forces SG2 and
Sk3, as well as for the different shapes oblate (o), prolate (p), or
spherical (s), where the minima occur for each isotope.
\par
\begin{tabular}{cccccccccc}
&  &  & SG2 &  &  &  &  & Sk3 &  \\ 
&  & 2qp & TDA & RPA &  &  & 2qp & TDA & RPA \\ 
&  &  &  &  &  &  &  &  &  \\ 
$^{64}$Ge & (o) & \multicolumn{1}{r}{14.2} & \multicolumn{1}{r}{14.2} & 
\multicolumn{1}{r}{9.8} &  & (o) & \multicolumn{1}{r}{14.3} & 
\multicolumn{1}{r}{14.3} & \multicolumn{1}{r}{9.9} \\ 
& (p) & \multicolumn{1}{r}{14.5} & \multicolumn{1}{r}{14.5} & 
\multicolumn{1}{r}{10.0} &  & (p) & \multicolumn{1}{r}{14.0} & 
\multicolumn{1}{r}{14.0} & \multicolumn{1}{r}{9.8} \\ 
$^{66}$Ge & (o) & \multicolumn{1}{r}{9.7} & \multicolumn{1}{r}{9.7} & 
\multicolumn{1}{r}{6.2} &  & (o) & \multicolumn{1}{r}{9.7} & 
\multicolumn{1}{r}{9.7} & \multicolumn{1}{r}{6.3} \\ 
& (p) & \multicolumn{1}{r}{10.6} & \multicolumn{1}{r}{10.6} & 
\multicolumn{1}{r}{6.7} &  & (p) & \multicolumn{1}{r}{10.1} & 
\multicolumn{1}{r}{10.1} & \multicolumn{1}{r}{6.5} \\ 
$^{68}$Ge & (o) & \multicolumn{1}{r}{7.4} & \multicolumn{1}{r}{7.4} & 
\multicolumn{1}{r}{4.3} &  & (o) & \multicolumn{1}{r}{7.4} & 
\multicolumn{1}{r}{7.4} & \multicolumn{1}{r}{4.3} \\ 
& (p) & \multicolumn{1}{r}{8.4} & \multicolumn{1}{r}{8.4} & 
\multicolumn{1}{r}{4.7} &  & (p) & \multicolumn{1}{r}{8.0} & 
\multicolumn{1}{r}{8.0} & \multicolumn{1}{r}{4.7} \\ 
$^{70}$Ge & (o) & \multicolumn{1}{r}{5.5} & \multicolumn{1}{r}{5.5} & 
\multicolumn{1}{r}{2.9} &  & (o) & \multicolumn{1}{r}{5.6} & 
\multicolumn{1}{r}{5.6} & \multicolumn{1}{r}{2.9} \\ 
& (p) & \multicolumn{1}{r}{6.6} & \multicolumn{1}{r}{6.6} & 
\multicolumn{1}{r}{3.2} &  & (p) & \multicolumn{1}{r}{6.7} & 
\multicolumn{1}{r}{6.7} & \multicolumn{1}{r}{3.5} \\ 
&  & \multicolumn{1}{r}{} & \multicolumn{1}{r}{} & \multicolumn{1}{r}{} &  & 
& \multicolumn{1}{r}{} & \multicolumn{1}{r}{} & \multicolumn{1}{r}{} \\ 
$^{68}$Se & (o) & \multicolumn{1}{r}{12.5} & \multicolumn{1}{r}{12.5} & 
\multicolumn{1}{r}{9.1} &  & (o) & \multicolumn{1}{r}{12.6} & 
\multicolumn{1}{r}{12.6} & \multicolumn{1}{r}{9.1} \\ 
& (p) & \multicolumn{1}{r}{13.5} & \multicolumn{1}{r}{13.5} & 
\multicolumn{1}{r}{9.6} &  & (p) & \multicolumn{1}{r}{13.2} & 
\multicolumn{1}{r}{13.2} & \multicolumn{1}{r}{9.5} \\ 
$^{70}$Se & (o) & \multicolumn{1}{r}{8.8} & \multicolumn{1}{r}{8.8} & 
\multicolumn{1}{r}{5.8} &  & (o) & \multicolumn{1}{r}{9.0} & 
\multicolumn{1}{r}{9.0} & \multicolumn{1}{r}{6.0} \\ 
& (p) & \multicolumn{1}{r}{10.0} & \multicolumn{1}{r}{10.0} & 
\multicolumn{1}{r}{6.3} &  & (p) & \multicolumn{1}{r}{9.8} & 
\multicolumn{1}{r}{9.8} & \multicolumn{1}{r}{6.4} \\ 
$^{72}$Se & (o) & \multicolumn{1}{r}{6.6} & \multicolumn{1}{r}{6.6} & 
\multicolumn{1}{r}{3.9} &  & (o) & \multicolumn{1}{r}{7.0} & 
\multicolumn{1}{r}{7.0} & \multicolumn{1}{r}{4.2} \\ 
& (p) & \multicolumn{1}{r}{7.7} & \multicolumn{1}{r}{7.7} & 
\multicolumn{1}{r}{4.2} &  & (p) & \multicolumn{1}{r}{9.7} & 
\multicolumn{1}{r}{9.7} & \multicolumn{1}{r}{5.6} \\ 
$^{74}$Se & (o) & \multicolumn{1}{r}{4.9} & \multicolumn{1}{r}{4.9} & 
\multicolumn{1}{r}{2.6} &  & (o) & \multicolumn{1}{r}{5.1} & 
\multicolumn{1}{r}{5.1} & \multicolumn{1}{r}{2.8} \\ 
& (s) & \multicolumn{1}{r}{4.7} & \multicolumn{1}{r}{4.7} & 
\multicolumn{1}{r}{2.4} &  & (p) & \multicolumn{1}{r}{8.1} & 
\multicolumn{1}{r}{8.1} & \multicolumn{1}{r}{4.3} \\ 
&  & \multicolumn{1}{r}{} & \multicolumn{1}{r}{} & \multicolumn{1}{r}{} &  & 
& \multicolumn{1}{r}{} & \multicolumn{1}{r}{} & \multicolumn{1}{r}{} \\ 
$^{72}$Kr & (o) & \multicolumn{1}{r}{10.9} & \multicolumn{1}{r}{10.9} & 
\multicolumn{1}{r}{8.0} &  & (o) & \multicolumn{1}{r}{11.3} & 
\multicolumn{1}{r}{11.3} & \multicolumn{1}{r}{8.3} \\ 
& (p) & \multicolumn{1}{r}{11.5} & \multicolumn{1}{r}{11.5} & 
\multicolumn{1}{r}{8.1} &  & (p) & \multicolumn{1}{r}{12.9} & 
\multicolumn{1}{r}{12.9} & \multicolumn{1}{r}{9.2} \\ 
$^{74}$Kr & (o) & \multicolumn{1}{r}{6.8} & \multicolumn{1}{r}{6.8} & 
\multicolumn{1}{r}{4.6} &  & (o) & \multicolumn{1}{r}{9.2} & 
\multicolumn{1}{r}{9.2} & \multicolumn{1}{r}{6.1} \\ 
& (p) & \multicolumn{1}{r}{11.3} & \multicolumn{1}{r}{11.3} & 
\multicolumn{1}{r}{7.3} &  & (p) & \multicolumn{1}{r}{11.5} & 
\multicolumn{1}{r}{11.5} & \multicolumn{1}{r}{7.5} \\ 
$^{76}$Kr & (s) & \multicolumn{1}{r}{4.7} & \multicolumn{1}{r}{4.7} & 
\multicolumn{1}{r}{2.8} &  & (o) & \multicolumn{1}{r}{6.2} & 
\multicolumn{1}{r}{6.2} & \multicolumn{1}{r}{3.7} \\ 
& (p) & \multicolumn{1}{r}{10.2} & \multicolumn{1}{r}{10.2} & 
\multicolumn{1}{r}{5.9} &  & (p) & \multicolumn{1}{r}{10.3} & 
\multicolumn{1}{r}{10.3} & \multicolumn{1}{r}{6.1} \\ 
$^{78}$Kr & (s) & \multicolumn{1}{r}{3.5} & \multicolumn{1}{r}{3.5} & 
\multicolumn{1}{r}{1.9} &  & (o) & \multicolumn{1}{r}{5.3} & 
\multicolumn{1}{r}{5.3} & \multicolumn{1}{r}{2.9} \\ 
&  & \multicolumn{1}{r}{} & \multicolumn{1}{r}{} & \multicolumn{1}{r}{} &  & 
(p) & \multicolumn{1}{r}{9.0} & \multicolumn{1}{r}{9.0} & \multicolumn{1}{r}{
4.8} \\ 
&  & \multicolumn{1}{r}{} & \multicolumn{1}{r}{} & \multicolumn{1}{r}{} &  & 
& \multicolumn{1}{r}{} & \multicolumn{1}{r}{} & \multicolumn{1}{r}{} \\ 
$^{76}$Sr & (o) & \multicolumn{1}{r}{8.7} & \multicolumn{1}{r}{8.7} & 
\multicolumn{1}{r}{6.6} &  & (o) & \multicolumn{1}{r}{9.4} & 
\multicolumn{1}{r}{9.4} & \multicolumn{1}{r}{7.0} \\ 
& (p) & \multicolumn{1}{r}{14.2} & \multicolumn{1}{r}{14.2} & 
\multicolumn{1}{r}{10.2} &  & (p) & \multicolumn{1}{r}{14.4} & 
\multicolumn{1}{r}{14.4} & \multicolumn{1}{r}{10.4} \\ 
$^{78}$Sr & (s) & \multicolumn{1}{r}{4.0} & \multicolumn{1}{r}{4.0} & 
\multicolumn{1}{r}{2.8} &  & (s) & \multicolumn{1}{r}{5.1} & 
\multicolumn{1}{r}{5.1} & \multicolumn{1}{r}{3.4} \\ 
& (p) & \multicolumn{1}{r}{12.1} & \multicolumn{1}{r}{12.1} & 
\multicolumn{1}{r}{7.8} &  & (p) & \multicolumn{1}{r}{12.4} & 
\multicolumn{1}{r}{12.3} & \multicolumn{1}{r}{8.1} \\ 
$^{80}$Sr & (s) & \multicolumn{1}{r}{3.7} & \multicolumn{1}{r}{3.7} & 
\multicolumn{1}{r}{2.4} &  & (s) & \multicolumn{1}{r}{4.8} & 
\multicolumn{1}{r}{4.8} & \multicolumn{1}{r}{2.9} \\ 
& (p) & \multicolumn{1}{r}{11.4} & \multicolumn{1}{r}{11.4} & 
\multicolumn{1}{r}{6.5} &  & (p) & \multicolumn{1}{r}{11.5} & 
\multicolumn{1}{r}{11.5} & \multicolumn{1}{r}{6.8} \\ 
$^{82}$Sr & (s) & \multicolumn{1}{r}{3.7} & \multicolumn{1}{r}{3.7} & 
\multicolumn{1}{r}{2.1} &  & (s) & \multicolumn{1}{r}{4.5} & 
\multicolumn{1}{r}{4.5} & \multicolumn{1}{r}{2.5}
\end{tabular}
\end{table}

\newpage

\begin{table}[tbp]
{\bf Table 6. }Same as Table 5 for the summed GT strength contained in the $%
Q_{EC}$ window $\left( \sum_{EC}\right) $.
\par
\begin{tabular}{ccccccccccc}
&  &  &  & SG2 &  &  &  &  & Sk3 &  \\ 
&  &  & 2qp & TDA & RPA &  &  & 2qp & TDA & RPA \\ 
&  &  &  &  &  &  &  &  &  &  \\ 
$^{64}$Ge &  & (o) & \multicolumn{1}{r}{10.7} & \multicolumn{1}{r}{1.2} & 
\multicolumn{1}{r}{1.0} &  & (o) & \multicolumn{1}{r}{9.5} & 
\multicolumn{1}{r}{1.2} & \multicolumn{1}{r}{1.0} \\ 
&  & (p) & \multicolumn{1}{r}{11.6} & \multicolumn{1}{r}{0.9} & 
\multicolumn{1}{r}{0.8} &  & (p) & \multicolumn{1}{r}{10.3} & 
\multicolumn{1}{r}{0.8} & \multicolumn{1}{r}{0.7} \\ 
$^{66}$Ge &  & (o) & \multicolumn{1}{r}{4.0} & \multicolumn{1}{r}{0.7} & 
\multicolumn{1}{r}{0.5} &  & (o) & \multicolumn{1}{r}{2.4} & 
\multicolumn{1}{r}{0.7} & \multicolumn{1}{r}{0.5} \\ 
&  & (p) & \multicolumn{1}{r}{6.4} & \multicolumn{1}{r}{0.5} & 
\multicolumn{1}{r}{0.4} &  & (p) & \multicolumn{1}{r}{1.9} & 
\multicolumn{1}{r}{0.5} & \multicolumn{1}{r}{0.3} \\ 
&  &  & \multicolumn{1}{r}{} & \multicolumn{1}{r}{} & \multicolumn{1}{r}{} & 
&  & \multicolumn{1}{r}{} & \multicolumn{1}{r}{} & \multicolumn{1}{r}{} \\ 
$^{68}$Se &  & (o) & \multicolumn{1}{r}{8.7} & \multicolumn{1}{r}{1.7} & 
\multicolumn{1}{r}{1.5} &  & (o) & \multicolumn{1}{r}{8.2} & 
\multicolumn{1}{r}{1.8} & \multicolumn{1}{r}{1.5} \\ 
&  & (p) & \multicolumn{1}{r}{10.1} & \multicolumn{1}{r}{2.0} & 
\multicolumn{1}{r}{1.8} &  & (p) & \multicolumn{1}{r}{9.2} & 
\multicolumn{1}{r}{1.6} & \multicolumn{1}{r}{1.4} \\ 
$^{70}$Se &  & (o) & \multicolumn{1}{r}{3.9} & \multicolumn{1}{r}{0.9} & 
\multicolumn{1}{r}{0.7} &  & (o) & \multicolumn{1}{r}{2.9} & 
\multicolumn{1}{r}{1.1} & \multicolumn{1}{r}{0.8} \\ 
&  & (p) & \multicolumn{1}{r}{6.2} & \multicolumn{1}{r}{1.5} & 
\multicolumn{1}{r}{1.3} &  & (p) & \multicolumn{1}{r}{6.0} & 
\multicolumn{1}{r}{1.6} & \multicolumn{1}{r}{0.4} \\ 
&  &  & \multicolumn{1}{r}{} & \multicolumn{1}{r}{} & \multicolumn{1}{r}{} & 
&  & \multicolumn{1}{r}{} & \multicolumn{1}{r}{} & \multicolumn{1}{r}{} \\ 
$^{72}$Kr &  & (o) & \multicolumn{1}{r}{7.6} & \multicolumn{1}{r}{3.4} & 
\multicolumn{1}{r}{2.7} &  & (o) & \multicolumn{1}{r}{7.0} & 
\multicolumn{1}{r}{3.0} & \multicolumn{1}{r}{2.3} \\ 
&  & (p) & \multicolumn{1}{r}{9.3} & \multicolumn{1}{r}{5.0} & 
\multicolumn{1}{r}{4.1} &  & (p) & \multicolumn{1}{r}{10.1} & 
\multicolumn{1}{r}{2.8} & \multicolumn{1}{r}{2.4} \\ 
$^{74}$Kr &  & (o) & \multicolumn{1}{r}{3.5} & \multicolumn{1}{r}{1.6} & 
\multicolumn{1}{r}{1.1} &  & (o) & \multicolumn{1}{r}{3.6} & 
\multicolumn{1}{r}{1.6} & \multicolumn{1}{r}{1.1} \\ 
&  & (p) & \multicolumn{1}{r}{7.3} & \multicolumn{1}{r}{2.3} & 
\multicolumn{1}{r}{1.9} &  & (p) & \multicolumn{1}{r}{7.3} & 
\multicolumn{1}{r}{1.7} & \multicolumn{1}{r}{1.5} \\ 
$^{76}$Kr &  & (s) & \multicolumn{1}{r}{1.6} & \multicolumn{1}{r}{0.6} & 
\multicolumn{1}{r}{0.3} &  & (o) & \multicolumn{1}{r}{1.9} & 
\multicolumn{1}{r}{0.9} & \multicolumn{1}{r}{0.5} \\ 
&  & (p) & \multicolumn{1}{r}{6.0} & \multicolumn{1}{r}{0.6} & 
\multicolumn{1}{r}{0.4} &  & (p) & \multicolumn{1}{r}{2.5} & 
\multicolumn{1}{r}{0.2} & \multicolumn{1}{r}{0.1} \\ 
&  &  & \multicolumn{1}{r}{} & \multicolumn{1}{r}{} & \multicolumn{1}{r}{} & 
&  & \multicolumn{1}{r}{} & \multicolumn{1}{r}{} & \multicolumn{1}{r}{} \\ 
$^{76}$Sr &  & (o) & \multicolumn{1}{r}{6.3} & \multicolumn{1}{r}{4.2} & 
\multicolumn{1}{r}{3.3} &  & (o) & \multicolumn{1}{r}{6.4} & 
\multicolumn{1}{r}{3.7} & \multicolumn{1}{r}{2.9} \\ 
&  & (p) & \multicolumn{1}{r}{12.1} & \multicolumn{1}{r}{5.2} & 
\multicolumn{1}{r}{4.7} &  & (p) & \multicolumn{1}{r}{12.2} & 
\multicolumn{1}{r}{3.1} & \multicolumn{1}{r}{3.0} \\ 
$^{78}$Sr &  & (s) & \multicolumn{1}{r}{2.0} & \multicolumn{1}{r}{1.4} & 
\multicolumn{1}{r}{0.9} &  & (s) & \multicolumn{1}{r}{2.9} & 
\multicolumn{1}{r}{2.2} & \multicolumn{1}{r}{1.3} \\ 
&  & (p) & \multicolumn{1}{r}{9.9} & \multicolumn{1}{r}{2.7} & 
\multicolumn{1}{r}{2.7} &  & (p) & \multicolumn{1}{r}{8.3} & 
\multicolumn{1}{r}{2.3} & \multicolumn{1}{r}{2.1} \\ 
$^{80}$Sr &  & (s) & \multicolumn{1}{r}{1.3} & \multicolumn{1}{r}{0.3} & 
\multicolumn{1}{r}{0.2} &  & (s) & \multicolumn{1}{r}{2.1} & 
\multicolumn{1}{r}{0.6} & \multicolumn{1}{r}{0.3} \\ 
&  & (p) & \multicolumn{1}{r}{7.1} & \multicolumn{1}{r}{1.0} & 
\multicolumn{1}{r}{0.9} &  & (p) & \multicolumn{1}{r}{6.3} & 
\multicolumn{1}{r}{0.8} & \multicolumn{1}{r}{0.5}
\end{tabular}
\end{table}

\newpage

\begin{table}[tbp]
{\bf Table 7. }Results from bare $2qp$, TDA, and RPA calculations for the
half-lives of Ge, Se, Kr, and Sr isotopes. The results correspond to the two
Skyrme forces SG2 and Sk3, as well as for the different shapes oblate (o),
prolate (p), or spherical (s), where the minima occur for each isotope.
Experimental values are from Ref. \cite{audi}.
\par
\begin{tabular}{cccccccccccc}
&  &  &  &  & SG2 &  &  &  &  & Sk3 &  \\ 
&  & exp. &  & 2qp & TDA & RPA &  &  & 2qp & TDA & RPA \\ 
&  &  &  &  &  &  &  &  &  &  &  \\ 
$^{64}$Ge & seconds & \multicolumn{1}{r}{63.7} & (o) & \multicolumn{1}{r}{14}
& \multicolumn{1}{r}{135} & \multicolumn{1}{r}{176} &  & (o) & 
\multicolumn{1}{r}{12} & \multicolumn{1}{r}{122} & \multicolumn{1}{r}{154}
\\ 
&  & \multicolumn{1}{r}{} & (p) & \multicolumn{1}{r}{20} & 
\multicolumn{1}{r}{208} & \multicolumn{1}{r}{259} &  & (p) & 
\multicolumn{1}{r}{19} & \multicolumn{1}{r}{161} & \multicolumn{1}{r}{206}
\\ 
$^{66}$Ge & hours & \multicolumn{1}{r}{2.26} & (o) & \multicolumn{1}{r}{0.3}
& \multicolumn{1}{r}{2.8} & \multicolumn{1}{r}{3.6} &  & (o) & 
\multicolumn{1}{r}{0.5} & \multicolumn{1}{r}{3.7} & \multicolumn{1}{r}{4.9}
\\ 
&  & \multicolumn{1}{r}{} & (p) & \multicolumn{1}{r}{0.2} & 
\multicolumn{1}{r}{2.9} & \multicolumn{1}{r}{4.1} &  & (p) & 
\multicolumn{1}{r}{0.6} & \multicolumn{1}{r}{4.9} & \multicolumn{1}{r}{7.2}
\\ 
&  & \multicolumn{1}{r}{} &  & \multicolumn{1}{r}{} & \multicolumn{1}{r}{} & 
\multicolumn{1}{r}{} &  &  & \multicolumn{1}{r}{} & \multicolumn{1}{r}{} & 
\multicolumn{1}{r}{} \\ 
$^{68}$Se & seconds & \multicolumn{1}{r}{35.5} & (o) & \multicolumn{1}{r}{18}
& \multicolumn{1}{r}{166} & \multicolumn{1}{r}{187} &  & (o) & 
\multicolumn{1}{r}{14} & \multicolumn{1}{r}{132} & \multicolumn{1}{r}{150}
\\ 
&  & \multicolumn{1}{r}{} & (p) & \multicolumn{1}{r}{12} & 
\multicolumn{1}{r}{111} & \multicolumn{1}{r}{128} &  & (p) & 
\multicolumn{1}{r}{14} & \multicolumn{1}{r}{105} & \multicolumn{1}{r}{129}
\\ 
$^{70}$Se & minutes & \multicolumn{1}{r}{41.1} & (o) & \multicolumn{1}{r}{12}
& \multicolumn{1}{r}{69} & \multicolumn{1}{r}{90} &  & (o) & 
\multicolumn{1}{r}{22} & \multicolumn{1}{r}{76} & \multicolumn{1}{r}{95} \\ 
&  & \multicolumn{1}{r}{} & (p) & \multicolumn{1}{r}{6} & \multicolumn{1}{r}{
39} & \multicolumn{1}{r}{44} &  & (p) & \multicolumn{1}{r}{19} & 
\multicolumn{1}{r}{137} & \multicolumn{1}{r}{180} \\ 
&  & \multicolumn{1}{r}{} &  & \multicolumn{1}{r}{} & \multicolumn{1}{r}{} & 
\multicolumn{1}{r}{} &  &  & \multicolumn{1}{r}{} & \multicolumn{1}{r}{} & 
\multicolumn{1}{r}{} \\ 
$^{72}$Kr & seconds & \multicolumn{1}{r}{17.2} & (o) & \multicolumn{1}{r}{3}
& \multicolumn{1}{r}{12} & \multicolumn{1}{r}{15} &  & (o) & 
\multicolumn{1}{r}{3} & \multicolumn{1}{r}{13} & \multicolumn{1}{r}{16} \\ 
&  & \multicolumn{1}{r}{} & (p) & \multicolumn{1}{r}{2} & \multicolumn{1}{r}{
11} & \multicolumn{1}{r}{13} &  & (p) & \multicolumn{1}{r}{3} & 
\multicolumn{1}{r}{15} & \multicolumn{1}{r}{21} \\ 
$^{74}$Kr & minutes & \multicolumn{1}{r}{11.5} & (o) & \multicolumn{1}{r}{1.1
} & \multicolumn{1}{r}{7.4} & \multicolumn{1}{r}{10.0} &  & (o) & 
\multicolumn{1}{r}{0.6} & \multicolumn{1}{r}{4.1} & \multicolumn{1}{r}{5.5}
\\ 
&  & \multicolumn{1}{r}{} & (p) & \multicolumn{1}{r}{2.3} & 
\multicolumn{1}{r}{10.4} & \multicolumn{1}{r}{14.0} &  & (p) & 
\multicolumn{1}{r}{3.1} & \multicolumn{1}{r}{19.4} & \multicolumn{1}{r}{28.7}
\\ 
$^{76}$Kr & hours & \multicolumn{1}{r}{14.8} & (s) & \multicolumn{1}{r}{0.9}
& \multicolumn{1}{r}{3.8} & \multicolumn{1}{r}{6.4} &  & (o) & 
\multicolumn{1}{r}{1.2} & \multicolumn{1}{r}{2.1} & \multicolumn{1}{r}{4.3}
\\ 
&  & \multicolumn{1}{r}{} & (p) & \multicolumn{1}{r}{0.2} & 
\multicolumn{1}{r}{2.3} & \multicolumn{1}{r}{3.1} &  & (p) & 
\multicolumn{1}{r}{0.7} & \multicolumn{1}{r}{12.1} & \multicolumn{1}{r}{15.0}
\\ 
&  & \multicolumn{1}{r}{} &  & \multicolumn{1}{r}{} & \multicolumn{1}{r}{} & 
\multicolumn{1}{r}{} &  &  & \multicolumn{1}{r}{} & \multicolumn{1}{r}{} & 
\multicolumn{1}{r}{} \\ 
$^{76}$Sr & seconds & \multicolumn{1}{r}{8.9} & (o) & \multicolumn{1}{r}{2}
& \multicolumn{1}{r}{9} & \multicolumn{1}{r}{11} &  & (o) & 
\multicolumn{1}{r}{2} & \multicolumn{1}{r}{9} & \multicolumn{1}{r}{11} \\ 
&  & \multicolumn{1}{r}{} & (p) & \multicolumn{1}{r}{4} & \multicolumn{1}{r}{
28} & \multicolumn{1}{r}{35} &  & (p) & \multicolumn{1}{r}{5} & 
\multicolumn{1}{r}{30} & \multicolumn{1}{r}{38} \\ 
$^{78}$Sr & minutes & \multicolumn{1}{r}{2.7} & (s) & \multicolumn{1}{r}{0.8}
& \multicolumn{1}{r}{2.3} & \multicolumn{1}{r}{3.3} &  & (s) & 
\multicolumn{1}{r}{0.3} & \multicolumn{1}{r}{0.7} & \multicolumn{1}{r}{1.1}
\\ 
&  & \multicolumn{1}{r}{} & (p) & \multicolumn{1}{r}{0.6} & 
\multicolumn{1}{r}{3.9} & \multicolumn{1}{r}{5.1} &  & (p) & 
\multicolumn{1}{r}{2.0} & \multicolumn{1}{r}{9.4} & \multicolumn{1}{r}{10.9}
\\ 
$^{80}$Sr & hours & \multicolumn{1}{r}{1.8} & (s) & \multicolumn{1}{r}{0.8}
& \multicolumn{1}{r}{4.4} & \multicolumn{1}{r}{6.8} &  & (s) & 
\multicolumn{1}{r}{0.7} & \multicolumn{1}{r}{3.1} & \multicolumn{1}{r}{5.0}
\\ 
&  & \multicolumn{1}{r}{} & (p) & \multicolumn{1}{r}{0.2} & 
\multicolumn{1}{r}{1.4} & \multicolumn{1}{r}{1.5} &  & (p) & 
\multicolumn{1}{r}{0.4} & \multicolumn{1}{r}{1.6} & \multicolumn{1}{r}{2.0}
\end{tabular}
\end{table}

\newpage

\begin{table}[tbp]
{\bf Table 8.} Results for various values of the particle-particle coupling
strength $\kappa _{pp}$ [MeV] from RPA calculations with the force SG2. The
table contains the summed GT strengths $\sum B_{GT}$ $\left[ g_{A}^{2}/4\pi
\right] $, the summed strengths up to $Q_{EC}$, $\sum_{EC}$ $\left[
g_{A}^{2}/4\pi \right] $, and the half-lives $T_{1/2}$ [minutes]. The
results are for $^{70}$Se with prolate and oblate shapes. The experimental
half-life is $T_{1/2}=41.1$ min.
\par
\begin{tabular}{ccccccccc}
&  & \multicolumn{3}{c}{prolate} &  & \multicolumn{3}{c}{oblate} \\ 
$\kappa _{pp}$ &  & $\sum B_{GT}$ & $\sum_{EC}$ & $T_{1/2}$ &  & $\sum
B_{GT} $ & $\sum_{EC}$ & $T_{1/2}$ \\ 
&  &  &  &  &  &  &  &  \\ 
0 &  & 6.33 & 1.31 & 43.9 &  & 5.81 & 0.67 & 89.6 \\ 
0.02 &  & 5.90 & 1.42 & 41.5 &  & 5.42 & 0.71 & 84.8 \\ 
0.05 &  & 5.15 & 1.69 & 35.3 &  & 4.74 & 0.80 & 74.0
\end{tabular}
\end{table}

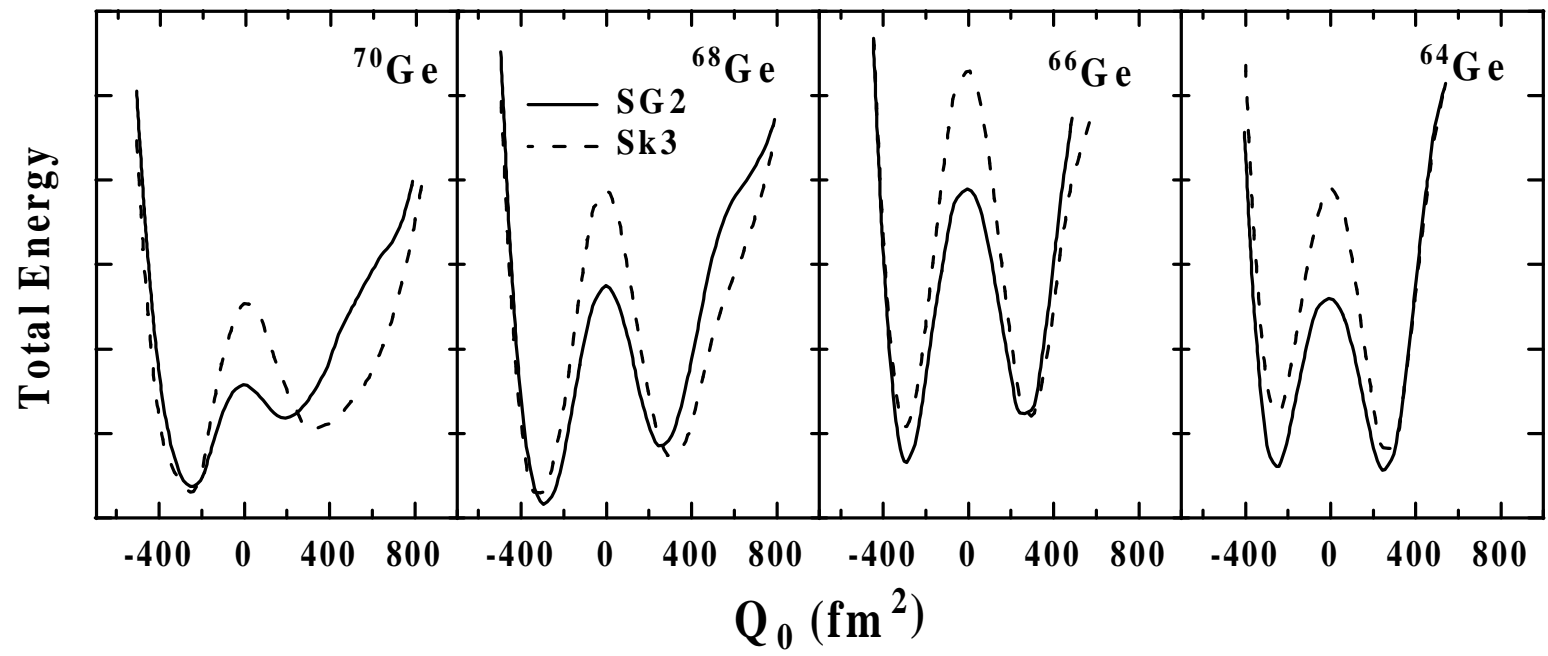
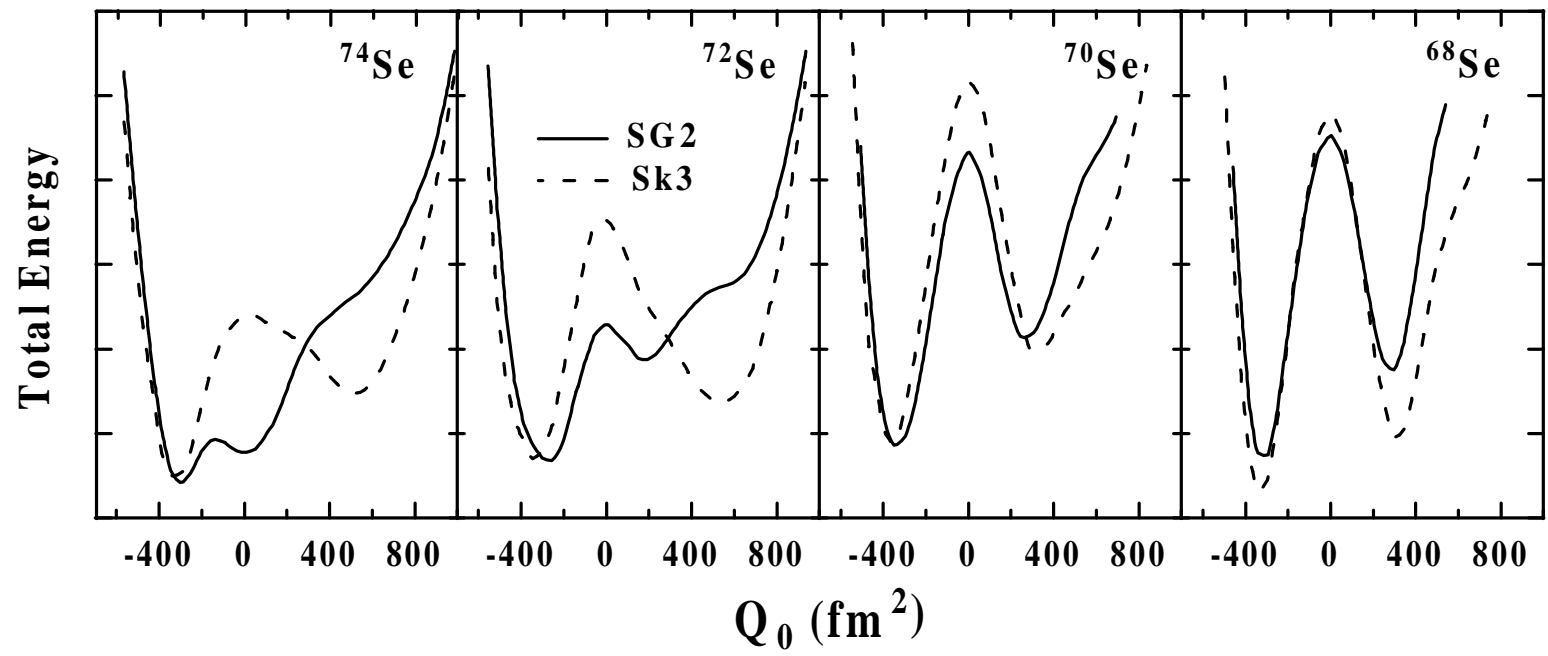
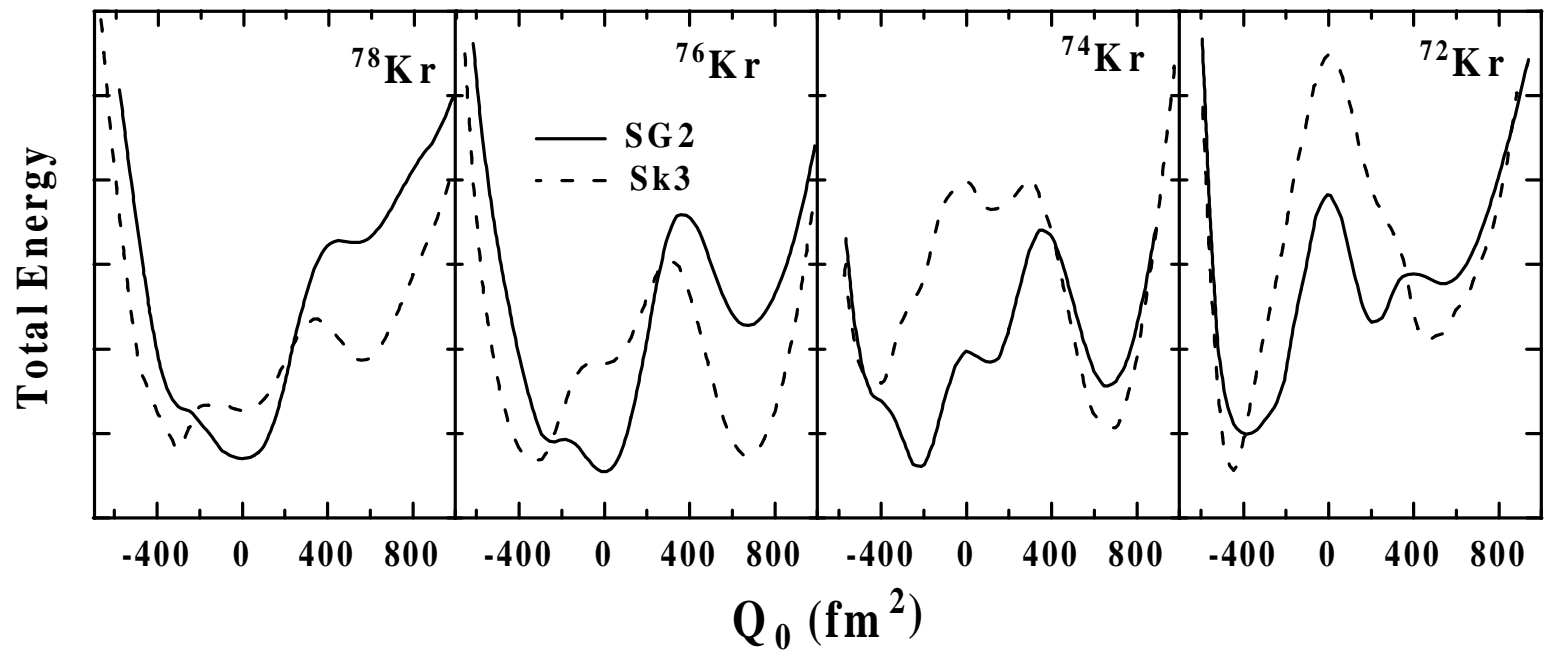
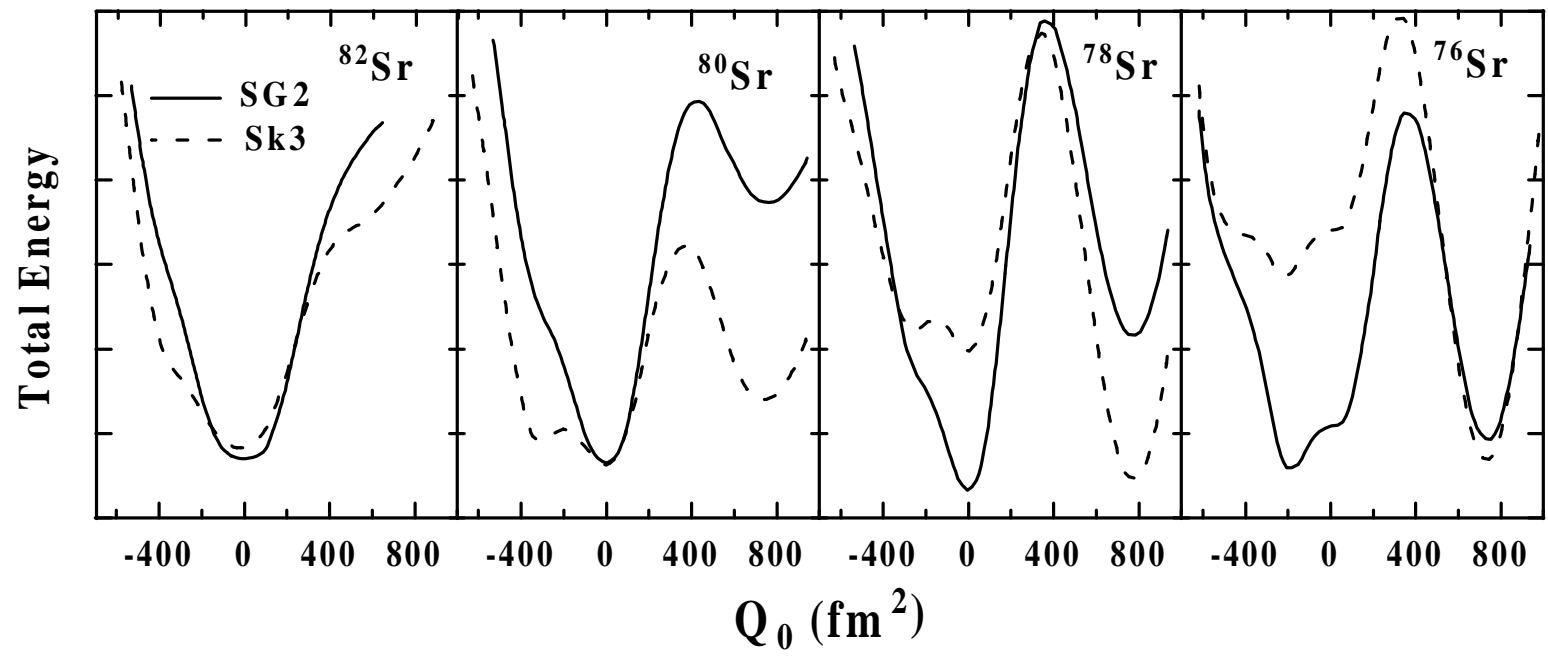
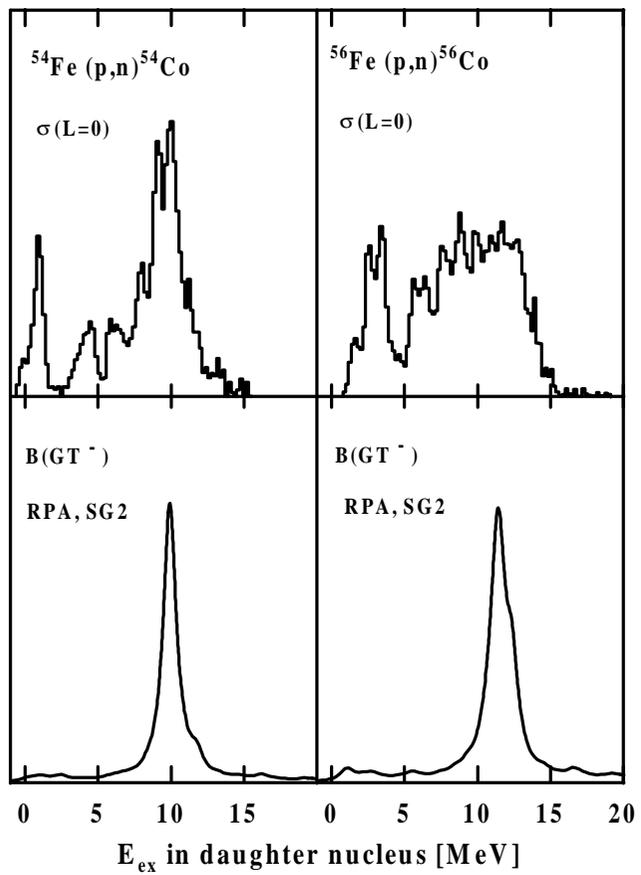
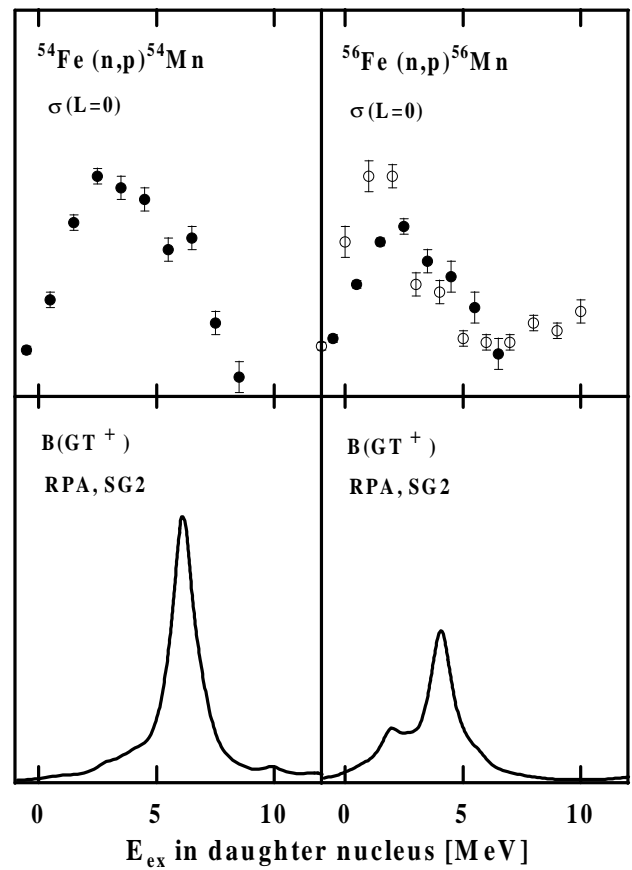
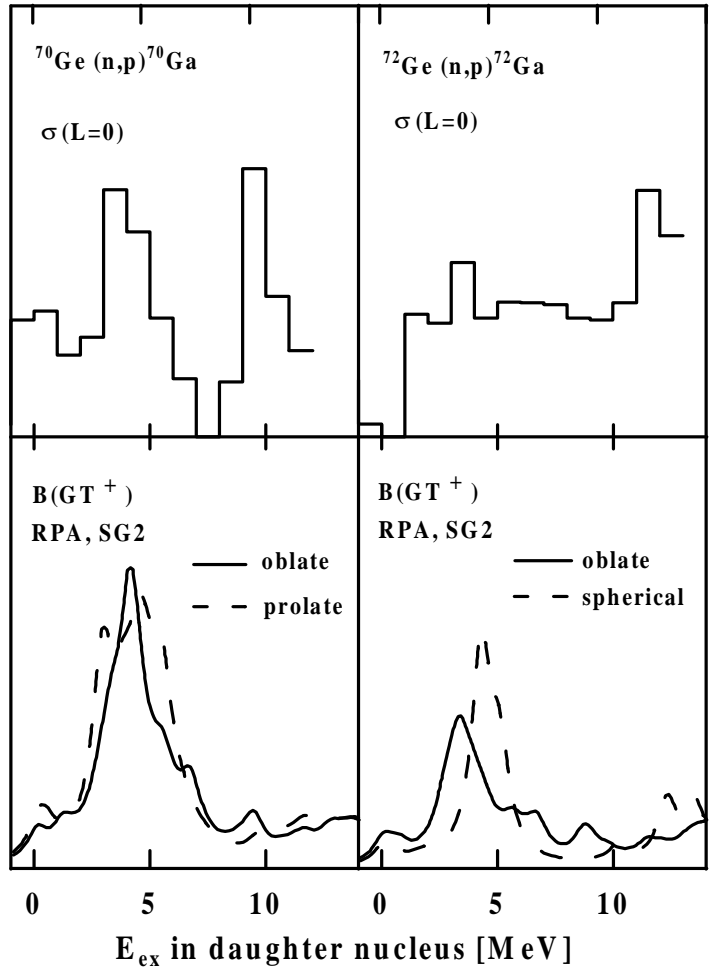
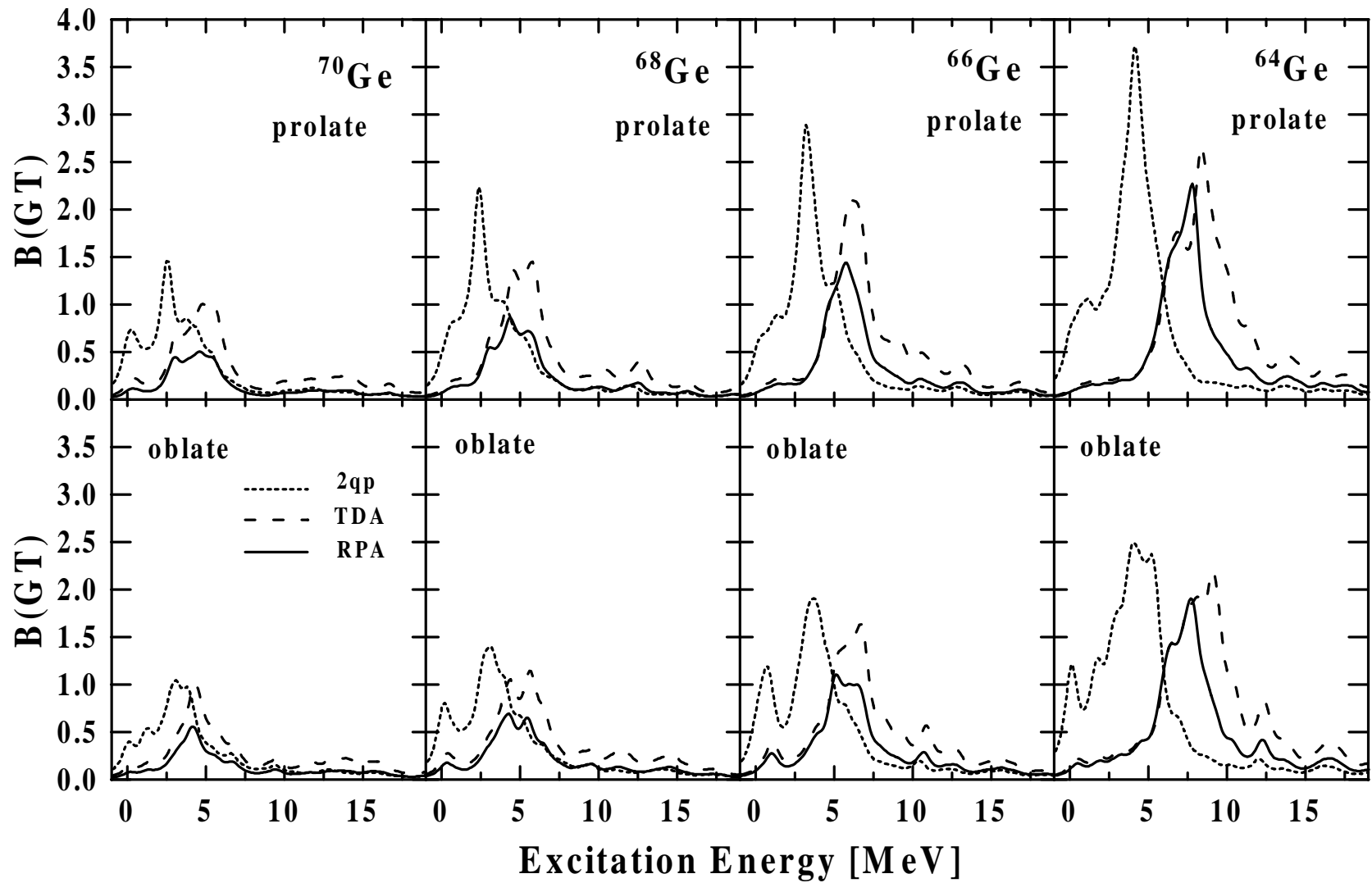
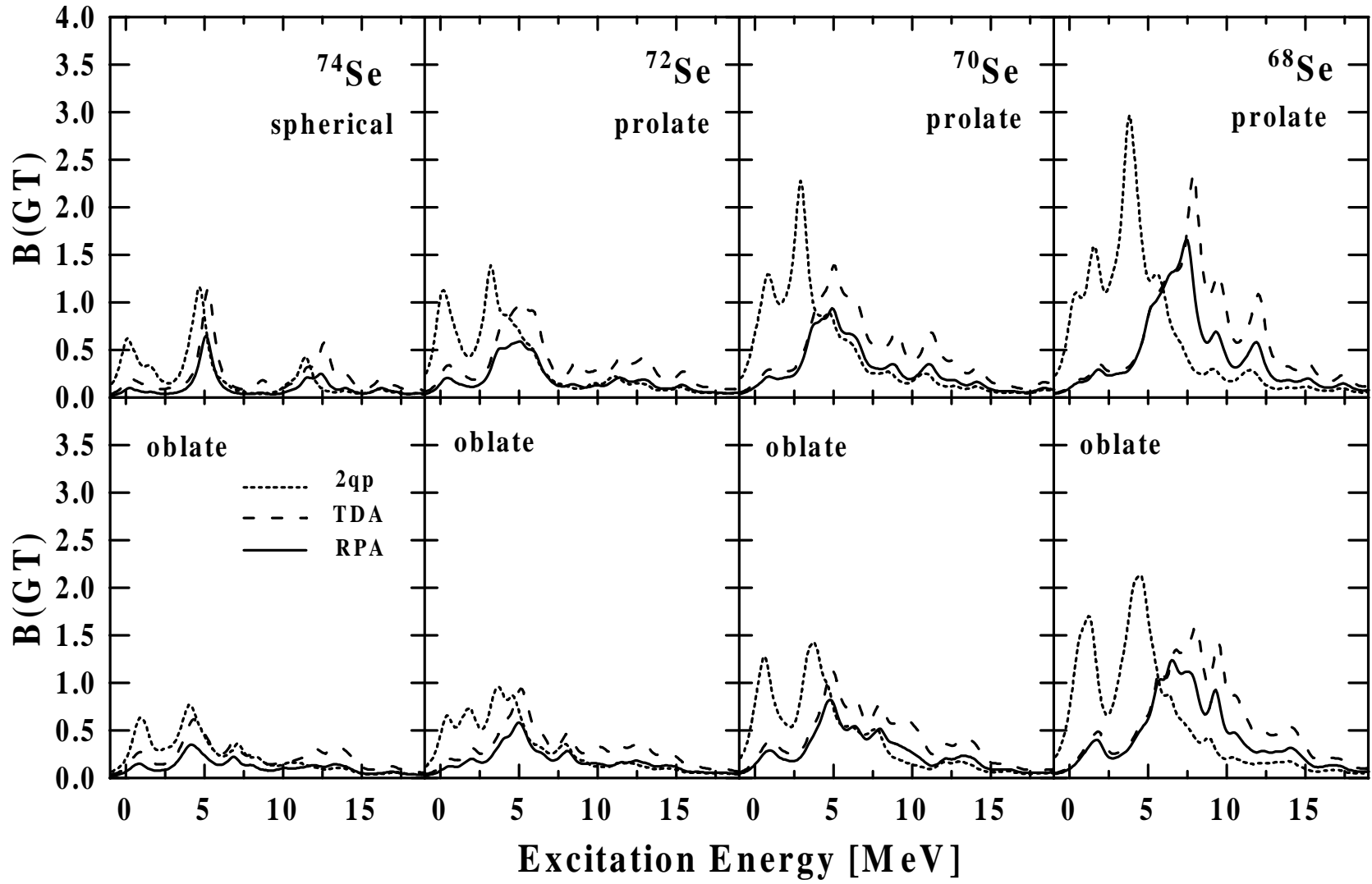
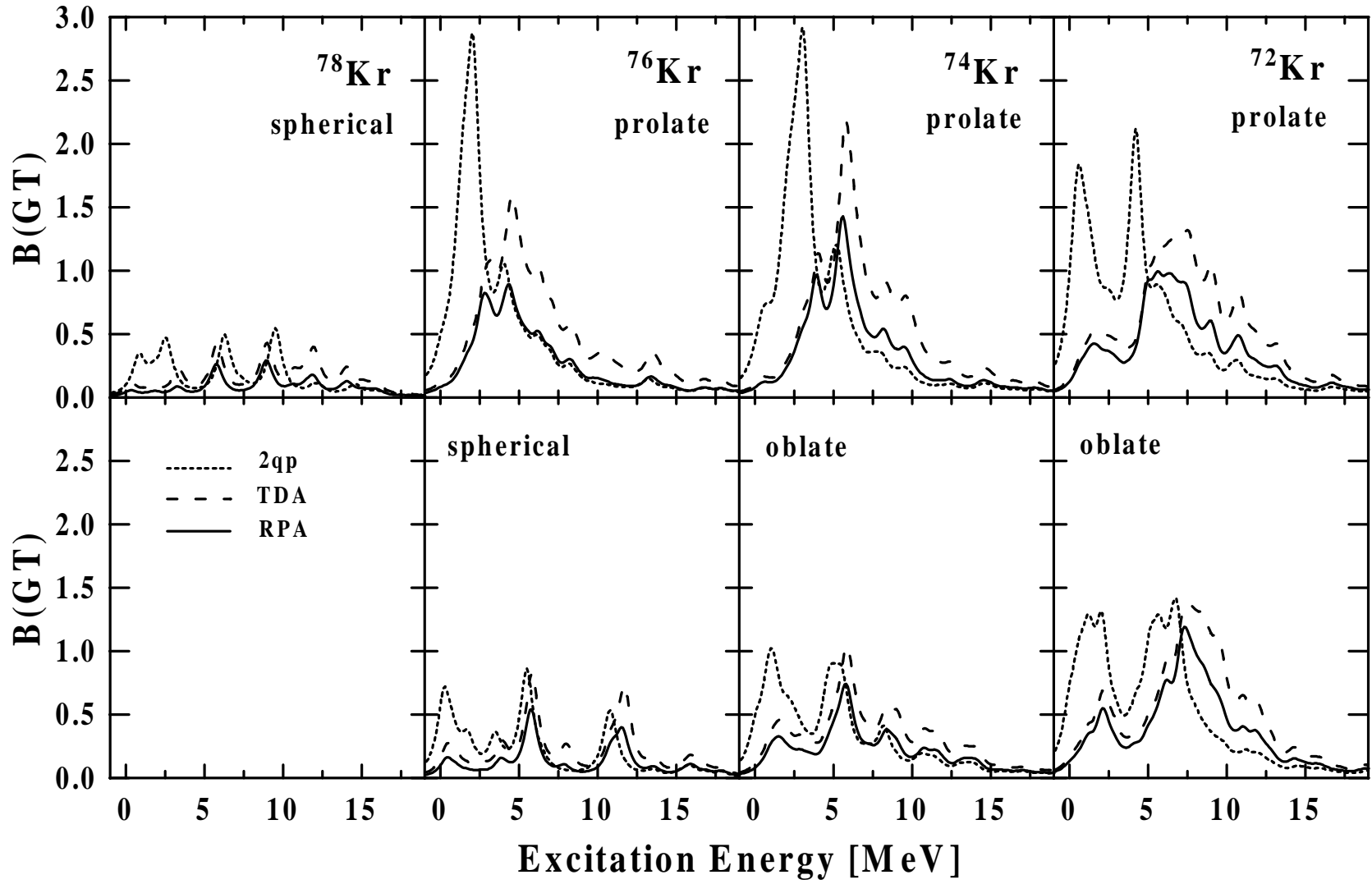
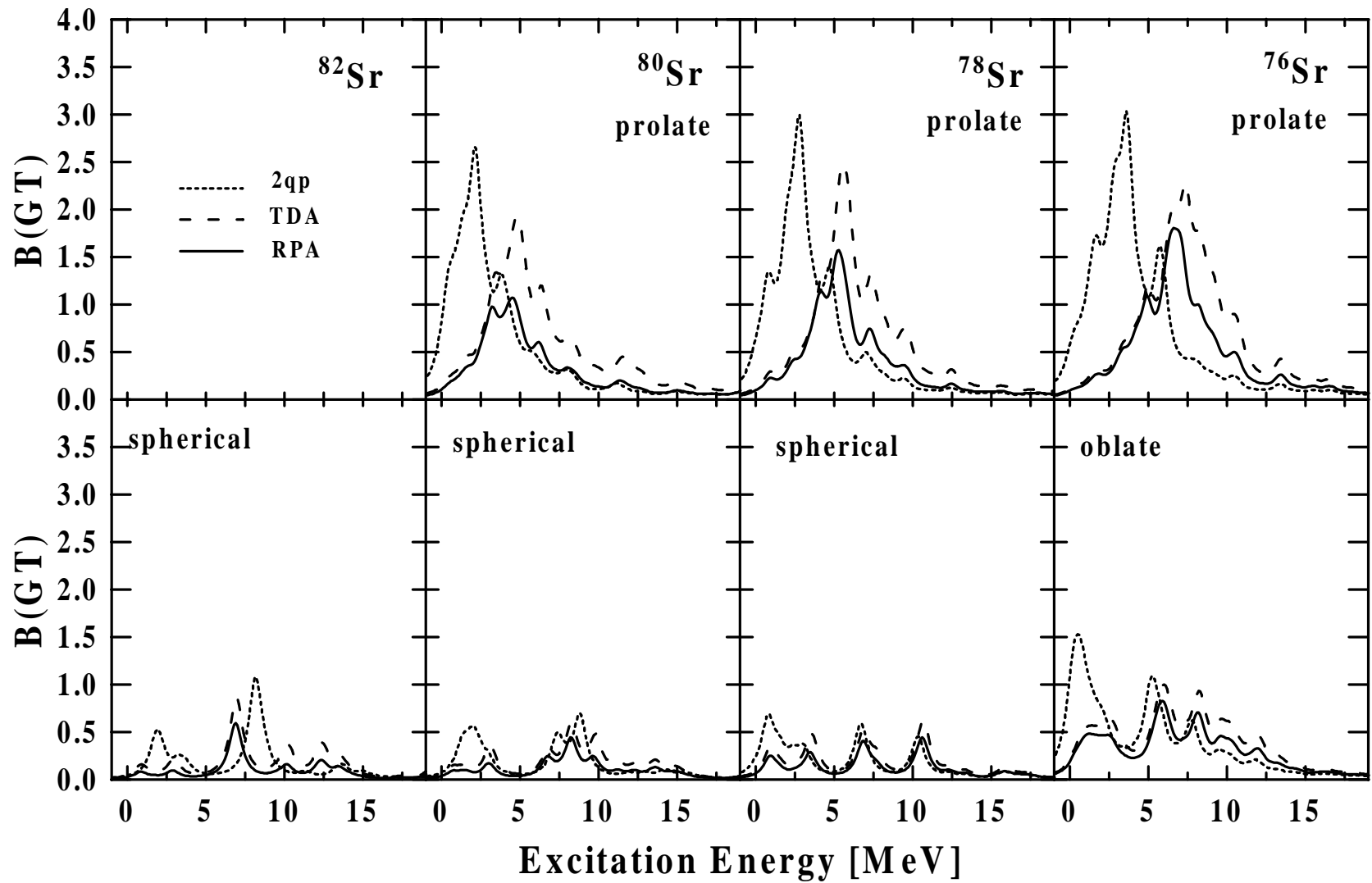
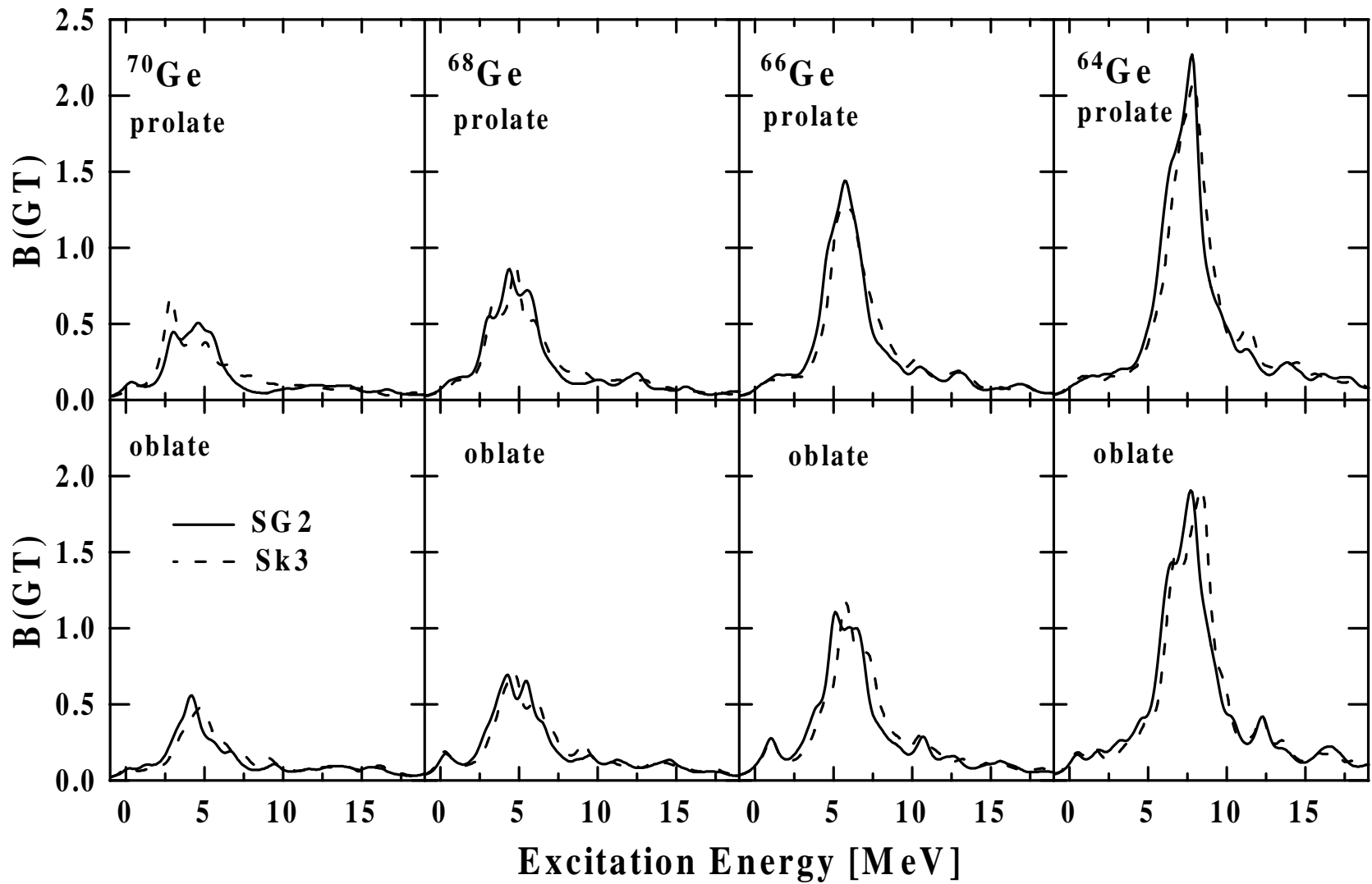
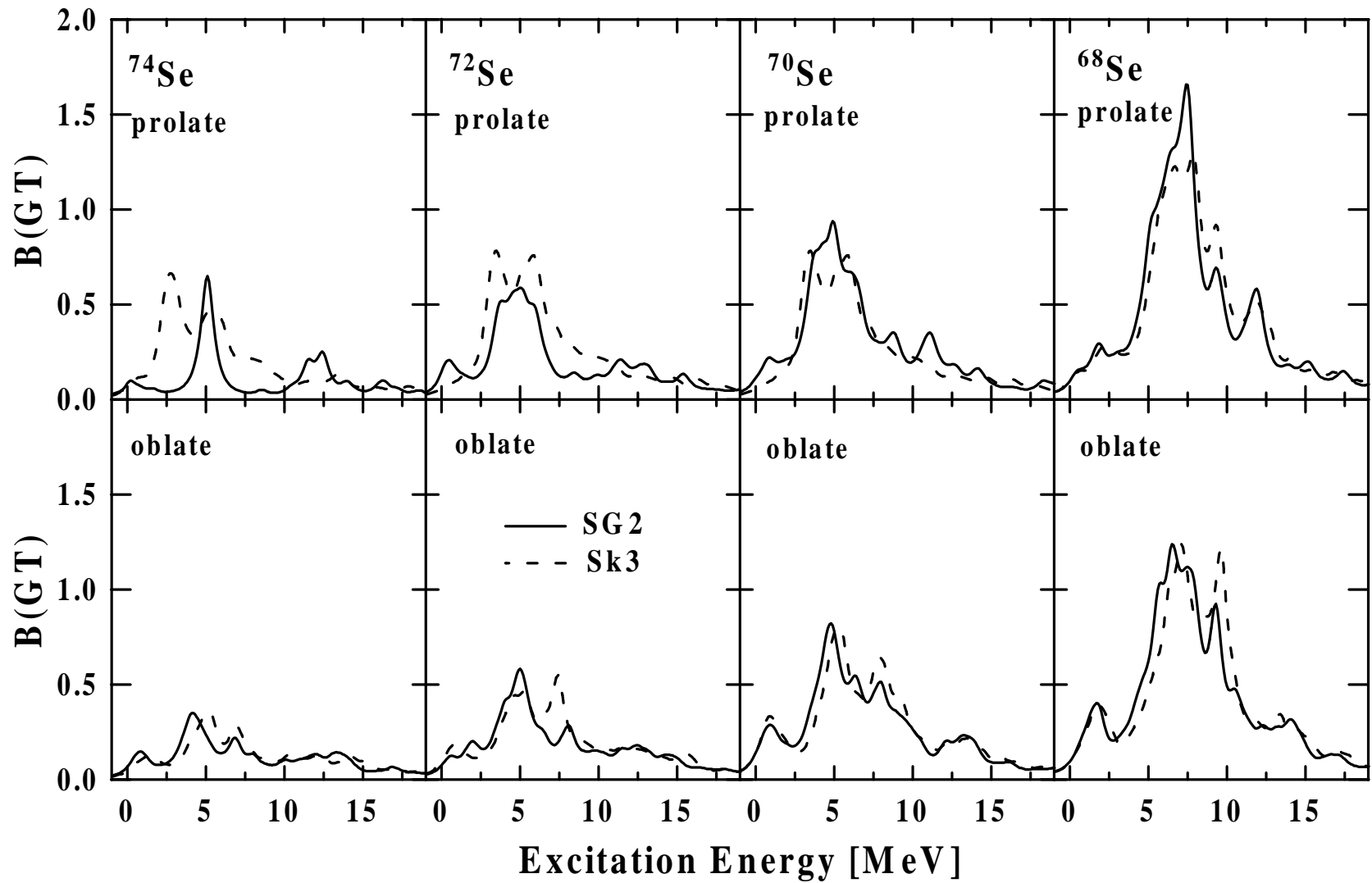
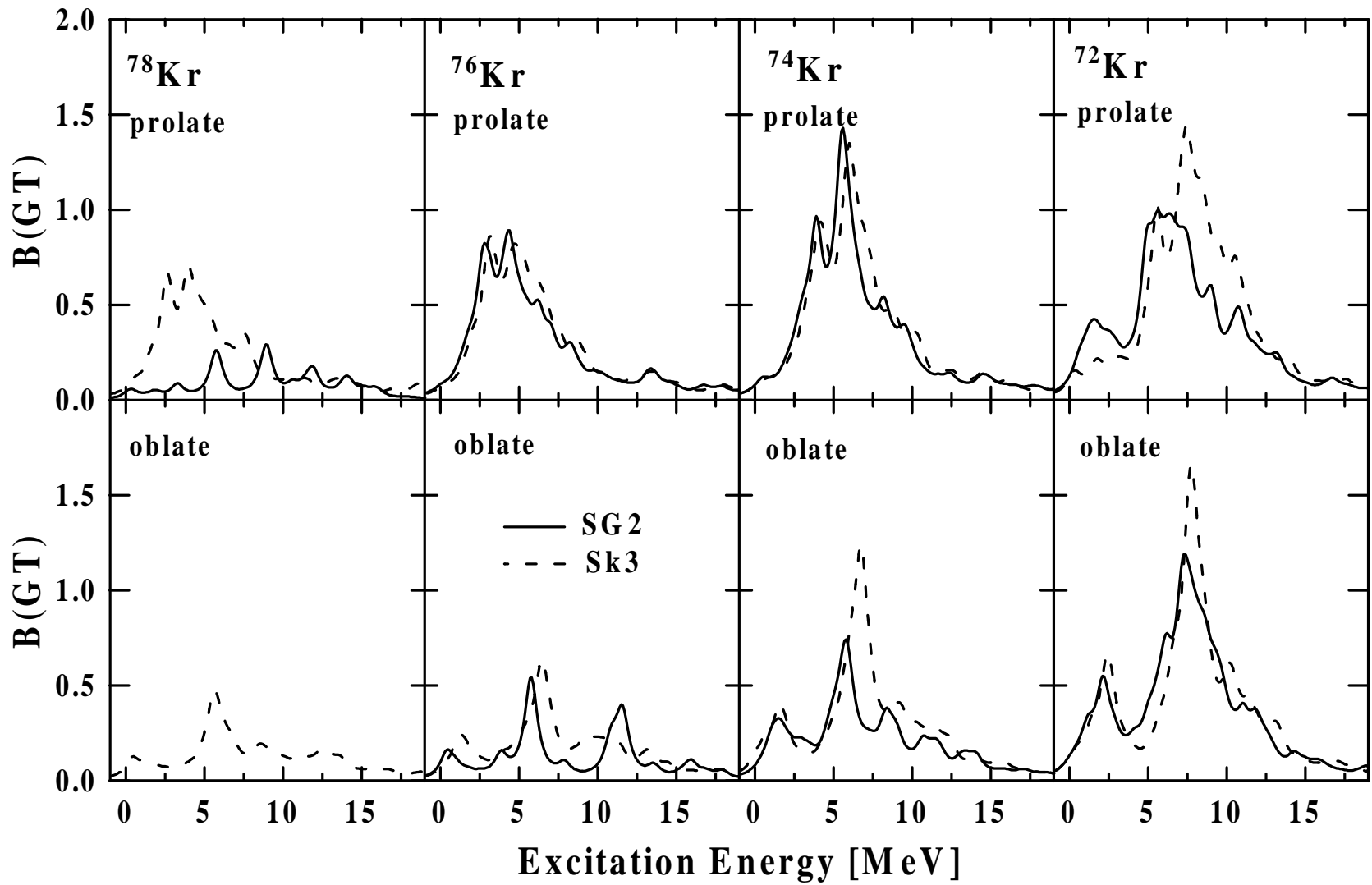
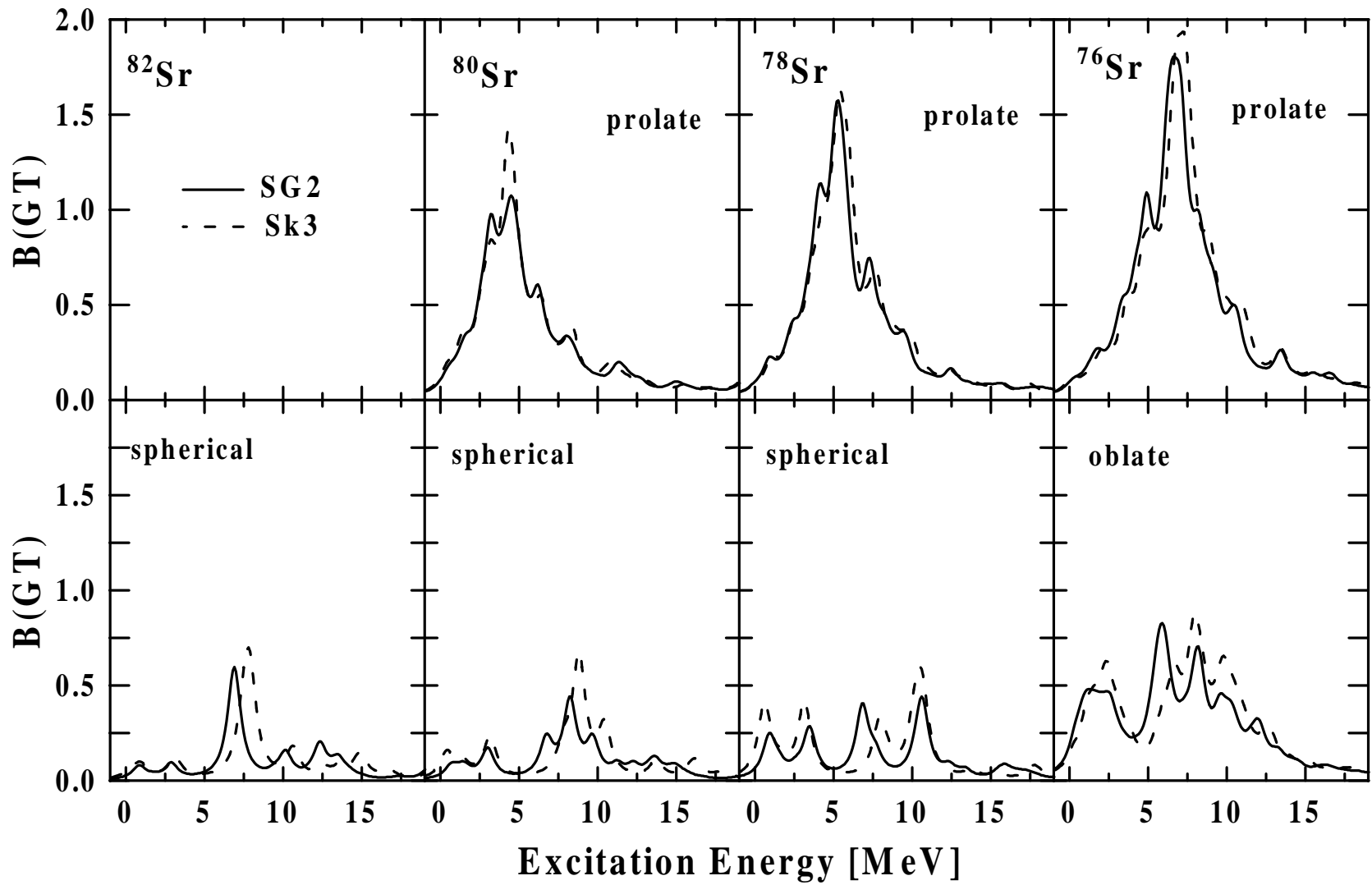
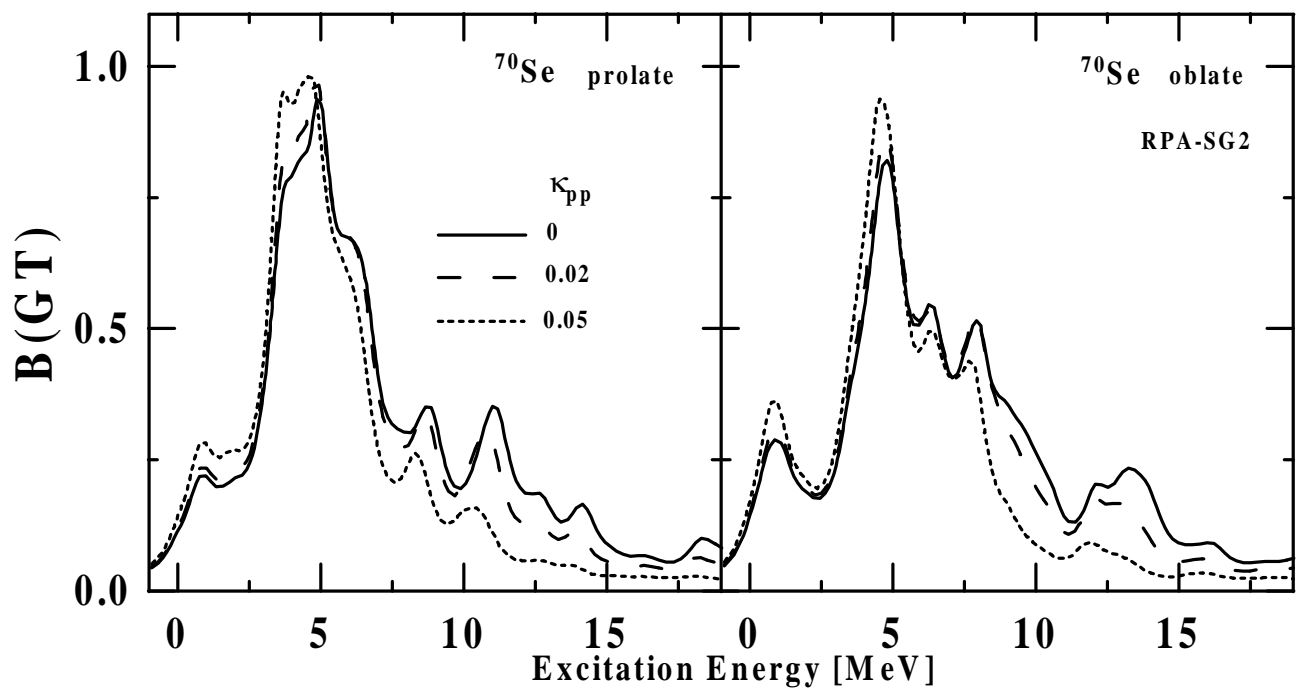
\end{document}